\documentclass[graybox, secnum]{svmult}

\usepackage{mathptmx}       
\usepackage{helvet}         
\usepackage{courier}        
\usepackage{type1cm}        
\usepackage{makeidx}         
\usepackage{graphicx}        
\usepackage{multicol}        
\usepackage[bottom]{footmisc}
\usepackage{hyperref}        
\usepackage{soul}            

\usepackage{amsmath}
\usepackage{amssymb}

\hypersetup{colorlinks=true,urlcolor=blue}
\usepackage[square,numbers]{natbib}
\makeindex             

\begin{document}
\title*{Tests of General Relativity using black hole X-ray data}
\author{Dimitry Ayzenberg \thanks{corresponding author} and Cosimo Bambi}
\institute{Dimitry Ayzenberg \at Theoretical Astrophysics, Eberhard-Karls Universit{\"a}t T{\"u}bingen, D-72076 T{\"u}bingen, Germany \\ \email{dimitry.ayzenberg@tat.uni-tuebingen.de}
\and Cosimo Bambi \at Center for Field Theory and Particle Physics and Department of Physics, Fudan University, 200438 Shanghai, China \\ \email{bambi@fudan.edu.cn}}
%
%
\maketitle
\abstract{The theory of General Relativity has successfully passed a large number of observational tests. The theory has been extensively tested in the weak-field regime with experiments in the Solar System and observations of binary pulsars. Recently, there have seen significant advancements in the study of the strong-field regime, which can now be tested with gravitational waves, X-ray data, and mm Very Long Baseline Interferometry observations. Here we summarize the state-of-the-art of the tests of General Relativity with black hole X-ray data, discussing its recent progress and future developments.}

\section*{Keywords} 
General Relativity; modified theories of gravity; black holes; Kerr metric; continuum-fitting method; iron line method.

\newpage 

\section{Introduction}

The theory of General Relativity was proposed by Einstein at the end of 1915 and has survived until today -- and without any modification -- as our standard framework for the description of the gravitational interaction and of the spacetime structure. The theory has been extensively tested in the so-called weak-field regime with experiments in the Solar System and radio observations of binary pulsars~\cite{Will:2014kxa}. The strong-field regime was almost completely unexplored up to some years ago, but recently there have been significant advancements. Black holes are ideal laboratories for testing General Relativity in the strong-field regime as these objects have the strongest gravitational fields that we can find in the Universe today~\cite{Bambi:2015kza,Yagi:2016jml,Cardoso:2016ryw}. Thanks to a new generation of observational facilities, we can now test black holes with gravitational waves, X-ray data, and mm Very Long Baseline Interferometry (VLBI) observations. 

Generally speaking, the analysis of black hole X-ray observations requires fitting the data with some theoretical model in order to measure the properties of the system. Some of these theoretical models rely on assumptions that may be violated in the presence of new physics. In General Relativity, we have~\cite{Will:2014kxa}:
\begin{enumerate}
\item The spacetime is described by the Kerr solution.
\item All particles follow the geodesics of the spacetime (Weak Equivalence Principle).
\item Atomic physics in the strong gravitational field of the black hole is the same as in our laboratories on Earth (Local Lorentz Invariance and Local Position Invariance).
\end{enumerate}
If the calculations of some spectral component have one or more of these assumptions, we can construct a theoretical model where such assumptions are regulated by one or more parameters (e.g. an assumption holds if a certain parameter vanishes and is violated if the parameter assumes a non-vanishing value). The analysis of the source can have such a parameter free and the fit can tell us if the data require a vanishing or non-vanishing parameter, namely the assumption holds or is violated. We note that, as of now, almost all tests of General Relativity with black hole X-ray data in the literature are devoted to test the Kerr metric, but there are also a few studies investigating geodesic motion~\cite{Roy:2021pns} and atomic physics~\cite{Bambi:2013mha} in the vicinity of black holes.

The aforementioned methodology is used not just for black hole X-ray tests, but also for other tests of gravity including gravitational wave tests. Black hole X-ray tests, and black hole electromagnetic tests in general, however, are different from gravitational wave tests in very important ways. Gravitational wave observations involve the merger of two compact objects and the detection of the perturbations on spacetime that propagate outwards from that merger. The compact objects involved are not solitary and the overall spacetime is not stationary, and so the spacetime cannot be described by the Kerr solution, but gravitational wave observations are the only ones that can probe the dynamical regime of gravity. Electromagnetic observations of black holes, on the other hand, involve black holes that are effectively in vacuum and stationary (any accretion disk is several orders of magnitude less massive and so does not significantly modify the spacetime~\cite{Bambi:2014koa}). One can then study the black hole spacetime and how matter interacts with the spacetime. Electromagnetic and gravitational wave tests can then be said to complement each other, as they probe different regimes of gravity that are not accessible to the other.

It is thus important to explore possible black hole electromagnetic tests as they can access a regime of gravity not available to other tests. Granted, there are still significant modeling uncertainties due to our lack of a perfect understanding of accretion disk physics, but this simply increases the importance of studying black hole accretion disks in order to reduce any uncertainties and be able to probe the spacetime more accurately. Indeed with the next generation of X-ray telescopes, such as eXTP and ATHENA (currently scheduled to be launched in 2027 and 2034, respectively)~\cite{eXTP:2016rzs,Nandra:2013jka}, it is now the time to more fully develop black hole accretion disk models particularly for testing General Relativity.

Along this vein, this chapter is meant as an overview of the necessary information for testing General Relativity using black hole X-ray data. We will focus on the needed background on black holes, accretion disk modeling, and modeling of X-ray observations for non-Kerr spacetimes. This chapter is broken up as follows: Section~\ref{sec:blackholes} gives an overview of black holes in General Relativity and beyond; Section~\ref{sec:disks} summarizes the thin disk model and departures from that model; Section~\ref{sec:tests} goes through the various black hole X-ray tests; summary and conclusions are in Section~\ref{sec:concs}.

\section{Black Holes}
\label{sec:blackholes}

\subsection{Black Holes in General Relativity}

Compared to other astrophysical objects, black holes are very simple, and this is especially true in General Relativity. As a result of the no-hair theorems, black holes in General Relativity are completely characterized by a small number of parameters under specific assumptions~\cite{1967PhRv..164.1776I, 1971PhRvL..26..331C, 1975PhRvL..34..905R, Chrusciel:2012jk}. In addition, the uniqueness theorems limit the possible black hole solutions to just a small family. These theorems can be circumvented by relaxing the assumptions or considering a theory beyond General Relativity, and is one of the main avenues of study in gravitational physics.

The simplest black hole solution in General Relativity is the Schwarzschild black hole. It describes a static, spherically-symmetric, and uncharged black hole that is completely desrcibed by just the black hole mass $M$. The equivalent charged solution is the Reissner-Nordstr{\"o}m black hole that has the additional parameter of electric charge $Q$. The Kerr solution describes a stationary, axisymmetric, and uncharged black hole with the parameters of black hole mass $M$ and spin angular momentum $J$. The general solution is the Kerr-Newman black hole that also includes the electric charge $Q$. Note that all of these solutions are asymptotically flat, i.e.~they become the Minkowski spacetime at spatial infinity, which is a reasonable assumption as we expect the gravitational influence of an object to vanish at spatial infinity.

Within astrophysics the Kerr black hole is primarily studied since macroscopic objects in the Universe are expected to have non-negligible angular momentum and negligible electric charge~\cite{Bambi:2017khi}. Note that the Schwarzschild solution is a special case of Kerr when the angular momentum vanishes. When studying black holes it is usually convenient to use geometric units where $G=c=1$, allowing us to write all parameters in units of the black hole's mass or radius. Along this vein, in place of the spin angular momentum $J$, a spin parameter $a$ or dimensionless spin parameter $\chi$ is usually used
\begin{equation}\label{eq-spin-chi}
    \chi = \frac{a}{M} = \frac{J}{M^{2}},
\end{equation}
or, if we restore $G$ and $c$, $\chi = cJ/GM^{2}$. In the literature, the notation $a_*$ to indicate the dimensionless spin parameter in Eq.~\ref{eq-spin-chi} is also often used.

The Kerr solution written as a spacetime line-element in the commonly used Boyer-Lindquist coordinates $(t,r,\theta,\phi)$ is~\cite{1963PhRvL..11..237K, Chandrasekhar:1985kt}
\begin{align}
    ds^{2} = & -\left(1-\frac{2Mr}{\Sigma}\right)dt^{2} - \frac{4Ma r\sin^{2}\theta}{\Sigma}dtd\theta + \frac{\Sigma}{\Delta}dr^{2}
    \nonumber \\
    & + \Sigma d\theta^{2} + \left(r^{2}+a^{2}+\frac{2Ma^{2}r\sin^{2}\theta}{\Sigma}\right)\sin^{2}\theta d\phi^{2},
\end{align}
where
\begin{equation}
    \Sigma\equiv r^{2}+a^{2}\cos^{2}\theta, \quad \Delta\equiv r^{2}-2Mr+a^{2}.
\end{equation}

Arguably the defining characteristic of a black hole is the event horizon. The horizon is defined as a null surface formed by marginally-trapped, null geodesics. The normal to this surface, $n^{a}$, must be null and so the event horizon must satisfy the horizon equation~\cite{Bambi:2017khi}
\begin{equation}
    g^{\alpha\beta}\partial_{\alpha}F\partial_{\beta}F=0, \label{eq:hor}
\end{equation}
where $g_{\alpha\beta}$ is the metric given by the line-element and $F(x^{\alpha})$ is a level surface function with normal $n_{\alpha}=\partial_{\alpha}F$. For any spacetime that is stationary, axisymmetric, and reflection symmetric about the poles and equator, as is the case for the Kerr solution, the level surfaces are only functions of radius $r$. Then, we can have $F(x^{\alpha})=r-r_{\text{hor}}$, where $r_{\text{hor}}$ is the event horizon radius. The horizon equation then reduces to $g^{rr}=0$ and for the Kerr solution the event horizon radius is
\begin{equation}
    r_{\text{hor}} = M\left(1+\sqrt{1-\chi^{2}}\right). \label{eq:khor}
\end{equation}

No null (e.g.~photons) and timelike (e.g.~massive particles) geodesics can leave from within the event horizon, and so the horizon is a terminal surface for astrophysics. Any photon or particle crossing the horizon is permanently gone and has no further impact on the system.

\subsection{Astrophysical Black Holes}

Astrophysical black holes are those that form from astrophysical processes, e.g.~stellar collapse or neutron star mergers, to contrast from primordial black holes that are theorized to form in the very early Universe~\cite{Carr:2005zd, 2010RAA....10..495K}. For the former, there is strong evidence of two classes of astrophysical black holes: stellar-mass black holes~\cite{Remillard:2006fc} and supermassive black holes~\cite{Kormendy:1995er}.
In addition, there is some evidence for intermediate-mass black holes that would fill the mass gap between stellar-mass and supermassive black holes~\cite{Miller:2003sc}. As astrophysical black holes form primarily from stellar collapse or mergers, there is a lower mass limit of $M\gtrsim 3~M_{\odot}$. Below this mass, the matter will be prevented from collapsing enough to form an event horizon due to the quantum pressure of neutrons and a neutron star will form instead~\cite{1974PhRvL..32..324R, 1996ApJ...470L..61K, 2012ARNPS..62..485L}.

\subsubsection{Stellar-mass Black Holes}

Stellar-mass black holes have a mass in the range $\sim3-\sim100~M_{\odot}$, i.e.~they have masses comparable to the masses of stars. When forming from stellar collapse, the mass of the remnant black hole is strongly dependent on the metallicity and mass of the progenitor star. As metallicity increases the mass loss rate through stellar wind increases, lowering the final mass of the progenitor when it collapses. The mechanism behind the collapse also varies depending on the metallicity and mass, further altering the mass of the remnant black hole, if forming one at all. For low metallicity stars~\cite{Heger:2001cd, Heger:2002by, 2015MNRAS.451.4086S}, there may be a mass gap for the remnant black hole between roughly $50$ and $150~M_{\odot}$. At higher metallicities, no remnant black holes above $150~M_{\odot}$ are expected as the mass loss rate is too high. Whether there are remnant black holes above $150~M_{\odot}$, even for low metallicity stars, is still debated as under some models the larger mass stars may undergo a runaway thermonuclear explosion that leaves no remnant~\cite{Heger:2001cd, Heger:2002by}.

Stellar evolution simulations suggest that there are $10^{8}-10^{9}$ stellar-mass black holes formed from stellar collapse in our Galaxy with similar numbers in similar galaxies~\cite{Timmes:1995kp}. All currently well-confirmed stellar-mass black holes are either in X-ray binaries or were detected through gravitational waves during a merger event. Those in X-ray binaries have masses in the range $M\sim3-20~M_{\odot}$ with, to date, about two dozen known with a dynamical measurement of the mass~\cite{Casares:2013tpa} and 50 more without a dynamical measurement of the mass~\cite{Corral-Santana:2015fud,blackcat}. Gravitational wave detectors, on the other hand, have detected about 250 stellar-mass black holes with masses ranging from the minimum of $M\sim 3~M_{\odot}$ up to $M\sim 175~M_{\odot}$~\cite{LIGOScientific:2021djp}. Note that the black holes detected through gravitational waves include at least one below the $3~M_{\odot}$ limit, but possibly above the maximum mass of a neutron star, and several in the $M\sim50-150~M_{\odot}$ formation mass gap pre-merger. In addition, it is worth noting that these merger events are one-time observations where two of the objects merge to become the final object, thus only about a third of the merger remnants now exist and are unlikely to be detected in additional mergers or through electromagnetic observations.

Here we are primarily interested in the stellar-mass black holes in X-ray binaries, and these can be grouped into two classes: low-mass X-ray binaries (LMXBs) and high-mass X-ray binaries (HMXBs). In LMXBs the companion star usually has a mass $M<3~M_{\odot}$, while in HMXBs the companion has a mass $M>10~M_{\odot}$. LMXBs tend to be transient X-ray sources, in that they are bright for a period of days to months and then quiescent for months to decades. This occurs because the mass transfer from the companion star is not constant, e.g.~the star can expand and overfill its Roche lobe, then gas will transfer to the black hole, and finally this gas transfer will contract the star back below its Roche lobe stopping the transfer. There are an expected $10^{3}-10^{4}$ LMXBs in the Galaxy~\cite{Yungelson:2006dn, Kiel:2006hd} and roughly 1-2 new ones are found each year when they pass from quiescence to an active state~\cite{blackcat}. HMXBs, on the other hand, tend to be persistent sources where the mass transfer is a relatively regular process, usually due to a stellar wind from the companion star. These systems are usually always a bright source without any quiescent periods.

\subsubsection{Supermassive Black Holes}

Supermassive black holes are those found at the centers of most galaxies. Most galaxies have a large density at their center, and for the galactic centers such as our own and of NGC 4258, there is strong evidence that these dense cores are black holes and not other objects~\cite{Maoz:1997yd}. While this cannot be so strongly confirmed for other galaxies, it is generally believed that all galaxies about the size of our own or larger have a supermassive black hole at their center. In the case of smaller galaxies, models and observations find differing conclusions and it is likely that some small galaxies do and some do not contain a supermassive black hole~\cite{2010A&ARv..18..279V, Volonteri:2007ax, Volonteri:2007dx, Ferrarese:2006fd, Gallo:2007xq}.

Supermassive black holes have masses of $M\sim 10^{5}-10^{10}~M_{\odot}$ and are thus the most massive single objects in the Universe. How they form, though, is still not well understood. The basic theory is that through accretion and mergers stellar-mass black holes can eventually grow to reach the higher masses of supermassive black holes. However, supermassive black holes with masses of $M\sim10^{10}~M_{\odot}$ have been observed in very distant galaxies when the Universe was less than $10^9$ years old~\cite{Madau:2014pta}. How these objects can reach such high masses so early is unclear~\cite{2010A&ARv..18..279V}. It may be the case that the original black holes form from the collapse of large clouds rather than stars, and so the initial masses are much larger, and/or these black holes experience a period of super-Eddington accretion~\cite{2010A&ARv..18..279V}.

\subsection{Black Holes Beyond General Relativity}
\label{sec:BHnonGR}

When studying black holes beyond General Relativity, particularly in the context of electromagnetic observations, generally two approaches are used: top-down and bottom-up. In the top-down approach we test the predictions of General Relativity, with the Kerr metric to describe the black hole spacetime, against another chosen theory of gravity in which black holes are not described by the Kerr metric. Two models must be constructed, one for the Kerr black hole in General Relativity and one for the black hole solution in the other theory. These models are then used to fit the data and it can be checked whether the data prefer one model over the other. Usually this leads to some constraint on the other theory of gravity rather than completely ruling out the theory. When one wishes to study a specific theory with a known rotating black hole solution, this is clearly the best approach. The difficulty with this approach arises because there are a large number of modified gravity theories and in very few of them have rotating black hole solutions been found (this is not surprising as it was almost 4 decades between the discovery of the Schwarzschild solution and the Kerr solution, and most of these modified gravity theories are more mathematically complex than General Relativity). Studying all of these various theories one-by-one is not a feasible pursuit.

The bottom-up approach, in contrast, is a phenomenological method where we wish to test General Relativity and/or the Kerr metric through a null experiment without considering any specific modified theory of gravity. In the case of electromagnetic observations, the spacetime can be described by a parametric black hole metric where additional deformation parameters are used to deform the Kerr metric. These deformation parameters quantify possible deviations from the Kerr solution and/or General Relativity and Kerr should be recovered when all deformation parameters vanish. While no specific theory is used to create the parametric metric it is possible that the metric can be mapped back to solutions in whatever theory one is interested. Regardless, the goal, as in any null-experiment, is to verify whether the Kerr hypothesis, our null hypothesis, is correct. Any non-vanishing deformation parameter required by astronomical data would violate the null hypothesis and suggest that the Kerr solution does not describe astrophysical black holes.

Although both methods have been used to study black hole X-ray observations, the bottom-up approach has been the most commonly used when analyzing actual data. There are several parametric black hole spacetimes available, but the most used to-date is known as the Johannsen metric~\cite{Johannsen:2013szh}. The Johannsen metric, which is a subset of a larger class of modified gravity metrics~\cite{Vigeland:2011ji}, introduces parameterized modifications that are motivated by a parameterized post-Newtonian (PPN) expansion~\cite{Will:2014kxa}, which is one very popular method for encoding departures from General Relativity.

Like the Kerr metric, the Johannsen metric is stationary, axisymmetric, and asymptotically-flat. It also has a Carter-like constant, an additional constant of the motion that exists in the Kerr metric, and so both metrics have similar symmetry properties. The line element of the Johannsen metric in Boyer-Lindquist coordinates $(t,r,\theta,\phi)$ reads
\begin{align}
    ds^{2} = & -\frac{-\tilde\Sigma\left(\Delta-a^{2}A_{2}^{2}\sin^{2}\theta\right)}{B^{2}}dt^{2} + \frac{\tilde\Sigma}{\Delta A_{5}}dr^{2} + \tilde\Sigma d\theta^{2}
    \nonumber \\
    & + \frac{\left[\left(r^{2}+a^{2}\right)^{2}A_{1}^{2}-a^{2}\Delta\sin^{2}\theta\right]\tilde\Sigma\sin^{2}\theta}{B^{2}}d\phi^{2}
    \nonumber \\
    & -\frac{2a\left[\left(r^{2}+a^{2}\right)A_{1}A_{2}-\Delta\right]\tilde\Sigma\sin^{2}\theta}{B^{2}}dtd\phi,
\end{align}
where
\begin{align}
    B = & \left(r^{2}+a^{2}\right)A_{1}-a^{2}A_{2}\sin^{2}\theta, \quad \tilde\Sigma=\Sigma+f,
    \nonumber \\
    \Sigma = & r^{2}+a^{2}\cos^{2}\theta, \quad \Delta=r^{2}-2Mr+a^{2},
\end{align}
the four free functions $f$, $A_{1}$, $A_{2}$, and $A_{5}$ are
\begin{align}
    f = & \sum^{\infty}_{n=3}\epsilon_{n}\frac{M^{n}}{r^{n-2}}, \quad A_{1} = 1+\sum^{\infty}_{n=3}\alpha_{1n}\left(\frac{M}{r}\right)^{n},
    \nonumber \\
    A_{2} = & 1+\sum^{\infty}_{n=2}\alpha_{2n}\left(\frac{M}{r}\right)^{n}, \quad A_{5} = 1+\sum^{\infty}_{n=2}\alpha_{5n}\left(\frac{M}{r}\right)^{n},
\end{align}
and as with the Kerr metric $M$ is the black hole mass and $a$ is the spin parameter of the black hole. There are additional lower-order terms in the four free functions, but these can be absorbed into the definition of $M$ and $a$ or must vanish to satisfy Solar System constraints. A more general extension of the Johannsen metric has been found~\cite{Carson:2020dez}, but has not yet been used for studying X-ray data. It should be noted that the deformation parameters of the Johannsen metric, or any other non-Kerr metric, have only been constrained with observational data individually, i.e.~with all but one set to zero. This is due to how computationally expensive it is to produce the models and analyze the data.

As with the Kerr solution it is important to know the location of the event horizon in the Johannsen metric. When solving the horizon equation, Eq.~\ref{eq:hor}, with the Johannsen metric we find that the event horizon radius matches that of Kerr, Eq.~\ref{eq:khor}. While the event horizon is at the same location other important surfaces, such as the innermost stable circular orbit (ISCO), are not the same and lead to significant modifications in the electromagnetic observations.

With any black hole spacetime it is important to study any pathologies that may be present, such as naked singularities. The Kerr spacetime, for example, has the requirement that the dimensionless spin parameter $|\chi| \leq 1$ so that an event horizon exists. The Johannsen metric and other non-Kerr metrics have additional requirements that are automatically satisfied in the Kerr solution, such as avoiding closed time-like curves and not violating the Lorentzian signature outside of the event horizon. Removing pathologies is often overlooked and can lead to unphysical results.

\section{Accretion Disks}
\label{sec:disks}

\subsection{Infinitesimally-thin Disks}
\label{sec:thindisks}

The standard accretion disk model that is used for black hole X-ray observations is the infinitesimally-thin disk model. It is a simplification of the Newtonian Shakura-Sunyaev~\cite{1973A&A....24..337S} and the relativistic Novikov-Thorne~\cite{1973blho.conf..343N} models that represent geometrically-thin and optically-thick disks. There is evidence the infinitesimally-thin disk model is a good approximation for accretion disks with accretion rates between $\sim5\%$ and $\sim30\%$ of the Eddington accretion rate~\cite{McClintock:2006xd,Steiner:2010kd,Kulkarni:2011cy}. Most work in analyzing black hole X-ray data also uses the infinitesimally-thin disk model for systems with much higher accretion rates, primarily because non-thin disk models are not in a form that is easily adapted for data analysis. There is recent work that has been done to alleviate this lack and will be discussed in the next subsection.

In the infinitesimally-thin disk model, the disk is confined to the equatorial plane of the black hole spacetime, i.e.~$\theta=\pi/2$ and $\dot\theta=0$, where the overhead dot represents a derivative with respect to proper time. Additionally, the disk is treated as stationary, i.e.~no explicit time-dependence, and the particles in the disk follow circular orbits (more realistically the particles follow quasi-circular orbits with decreasing radius, but the timescale of the radial motion is larger than the orbital timescale and so can usually be ignored). The far majority of black hole spacetimes that are studied in the context of black hole X-ray observations are stationary and axisymmetric, and so they possess a timelike Killing vector and an azimuthal Killing vector. These Killing vectors are associated with the conservation of the specific energy $E$ and the $z$-component of the specific angular momentum $L_{z}$ of the particles in the disk. These conserved quantities and the prior imposed conditions make the black hole-thin disk system fully determined~\cite{1972ApJ...178..347B,Bambi:2017khi}.

From the definitions of $E$ and $L_{z}$ we get
\begin{align}
    \dot t = & -\frac{Eg_{\phi\phi}+L_{z}g_{t\phi}}{g_{tt}g_{\phi\phi}-g_{t\phi}^{2}},
    \\
    \dot\phi = & \frac{Eg_{t\phi}+L_{z}g_{tt}}{g_{tt}g_{\phi\phi}-g_{t\phi}^{2}},
\end{align}
where the overhead dot is again a derivative with respect to proper time. By substituting the above into the normalization condition for the four-velocity of massive particles, $u^{\alpha}u_{\alpha}=-1$, we find
\begin{equation}
    g_{rr}\dot r^{2}+g_{\theta\theta}\dot\theta^{2} = V_{\text{eff}}\left(r,\theta;E,L_{z}\right),
\end{equation}
where the effective potential is
\begin{equation}
    V_{\text{eff}} = -1-\frac{E^{2}g_{\phi\phi}+2EL_{z}g_{t\phi}+L_{z}^{2}g_{tt}}{g_{tt}g_{\phi\phi}-g_{t\phi}^{2}}, \label{eq:Veff}
\end{equation}
and the four-velocity is $u^{\alpha}=\left(\dot t, \dot r, \dot\theta, \dot\phi\right)$.

Restricting to equatorial and circular orbits, where $\dot{r}=\dot{\theta}=\ddot{r}=0$, explicit expressions for the energy and angular momentum can be obtained~\cite{Bambi:2017khi}
\begin{align}
    E = & -\left(g_{tt}+g_{t\phi}\Omega\right)\dot t = -\frac{g_{tt}+g_{t\phi}\Omega}{\sqrt{-\left(g_{tt}+2g_{t\phi}\Omega+g_{\phi\phi}\Omega^{2}\right)}}, \label{eq:E}
    \\
    L_{Z} = & \left(g_{t\phi}+g_{\phi\phi}\Omega\right)\dot t = \frac{g_{t\phi}+g_{\phi\phi}\Omega}{\sqrt{-\left(g_{tt}+2g_{t\phi}\Omega+g_{\phi\phi}\Omega^{2}\right)}}, \label{eq:Lz}
\end{align}
where the angular velocity of the equatorial circular geodesics is
\begin{equation}
    \Omega = \frac{d\phi}{dt} = \frac{-g_{t\phi,r}\pm\sqrt{\left(g_{t\phi,r}\right)^{2}-g_{tt,r}g_{\phi\phi,r}}}{g_{\phi\phi,r}}, \label{eq:omega}
\end{equation}
and
\begin{equation}
    \dot t = \frac{1}{\sqrt{-\left(g_{tt}+2g_{t\phi}\Omega+g_{\phi\phi}\Omega^{2}\right)}}. \label{eq:dt}
\end{equation}

For black hole X-ray observations the necessary quantity to calculate from the accretion disk model is the redshift experienced by photons emitted from the disk. The redshift factor is the ratio of the observed frequency to the emitted frequency and is given by
\begin{equation}
    g = \frac{\nu_{o}}{\nu_{e}} = \frac{\left(p_{\alpha}u^{\alpha}\right)_{o}}{\left(p_{\beta}u^{\beta}\right)_{e}}.
\end{equation}
Here $p_{\alpha}$ is the canonical conjugate momentum of a photon traveling from the emitter to the observer, and $u^{\alpha}_{o}$ and $u^{\alpha}_{e}$ are the four-velocities of the observer and emitter, respectively.

By sticking to stationary and axisymmetric spacetimes, the photon's conjugate momentum can be written $p_{\alpha}=\left(-E^{\gamma},p_{r},p_{\theta},L_{z}^{\gamma}\right)$. The observer can be treated as static, $u^{\alpha}_{o}=\left(1,0,0,0\right)$, and the four-velocity of the orbiting emitting material has already been calculated, $u^{\alpha}_{e}=u^{t}_{e}\left(1,0,0,\Omega\right)$. Here $u^{t}_{e}=\dot t$, which is given by Eq.~\ref{eq:dt}, and $\Omega$ is given by Eq.~\ref{eq:omega}. The redshift factor can now be shown to be
\begin{equation}
    g = \frac{\sqrt{-\left(g_{tt}+2g_{t\phi}\Omega+g_{\phi\phi}\Omega^{2}\right)}}{1-\Omega b}, \label{eq:red}
\end{equation}
where $b\equiv L_{z}^{\gamma}/E^{\gamma}$ is a conserved quantity usually called the impact parameter.

Some models also make use of the emission angle $\theta_{e}$, which is necessary if the local emission of the disk is not isotropic. The normal to the disk is given by
\begin{equation}
    n^{\alpha} = \left(0,0,\sqrt{g^{\theta\theta}},0\right)|_{r_{e},\theta_{e}=\pi/2},
\end{equation}
and so the emission angle is
\begin{equation}
    \cos\theta_{e} = \frac{n^{\alpha}p_{\alpha}}{u^{\beta}_{e}p_{\beta}}|_{e}=g\sqrt{g^{\theta\theta}}\frac{p^{e}_{\theta}}{p^{e}_{t}},
\end{equation}
where $p^{e}_{\alpha}$ is the photon conjugate momentum at the emission point in the disk.

Finally, the ISCO for massive particles in the disk can also be calculated. Circular orbits for massive particles are unstable inside the ISCO, and so they are expected to quickly plunge and cross the event horizon. Most models assume that the inner edge of the accretion disk cannot be smaller than the ISCO radius, and in even simpler cases set the inner edge to the ISCO radius. As the ISCO is the boundary between stable and unstable orbits one can calculate the ISCO by solving $\partial^{2}V_{\text{eff}}/\partial r^{2}=0$ or $\partial^{2}V_{\text{eff}}/\partial \theta^{2}=0$ after substituting the expressions for energy and angular momentum, Eqs.~\ref{eq:E} and~\ref{eq:Lz} respectively, into the expression for the effective potential, Eq.~\ref{eq:Veff}\footnote{We note that, in the Kerr metric and in many non-Kerr metrics, equatorial circular orbits are always vertically stable and therefore the ISCO radius is determined by the condition $\partial^{2}V_{\text{eff}}/\partial r^{2}=0$. However, in some non-Kerr spacetimes the ISCO may be determined by the stability of the orbit along the vertical direction~\cite{Bambi:2011vc} and/or there may be annuli of unstable orbits between annuli of stable orbits~\cite{Bambi:2013eb}.}.

\begin{figure*}[hpt]
\includegraphics[width=0.33\columnwidth{},clip=true]{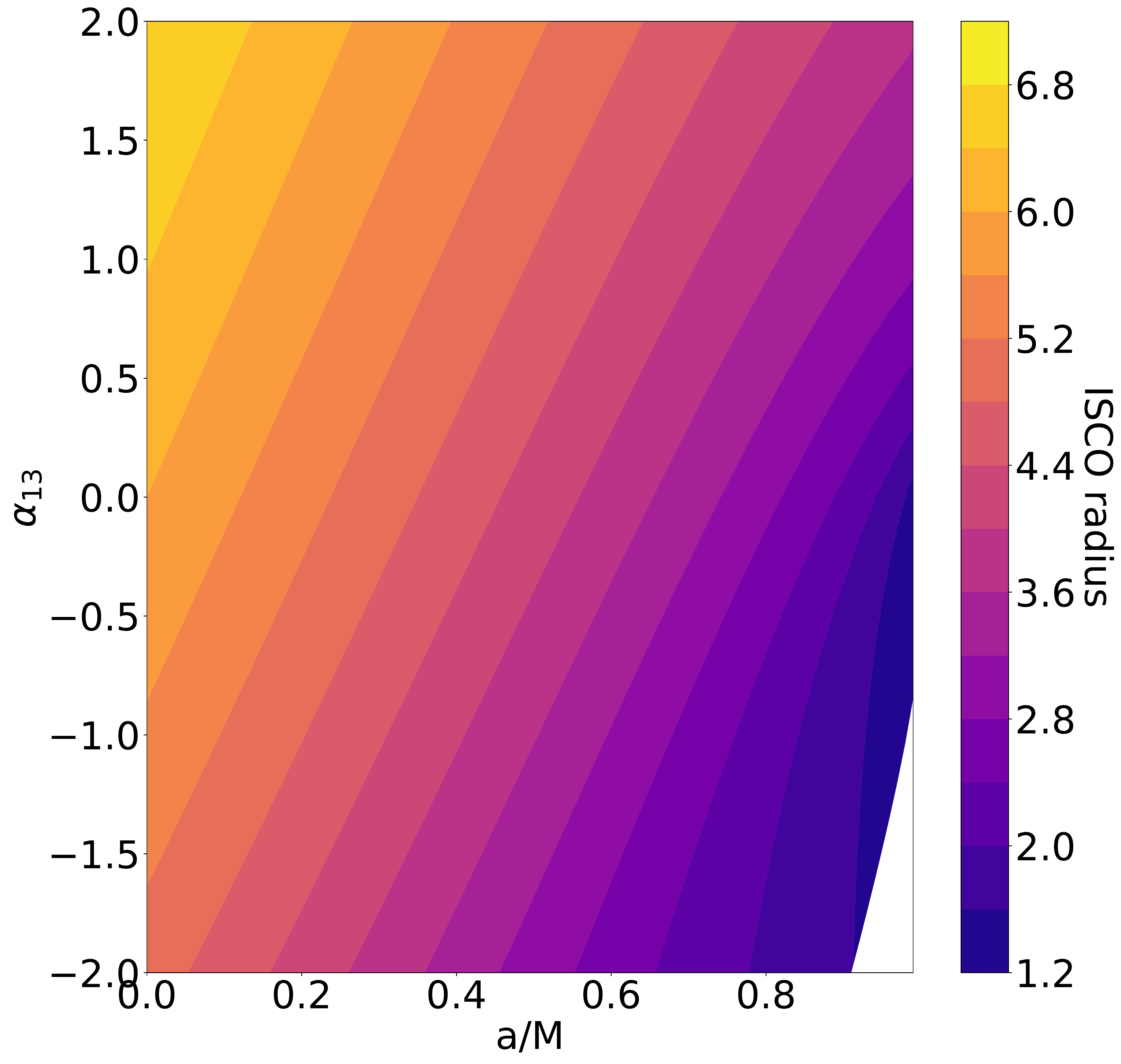}
\includegraphics[width=0.33\columnwidth{},clip=true]{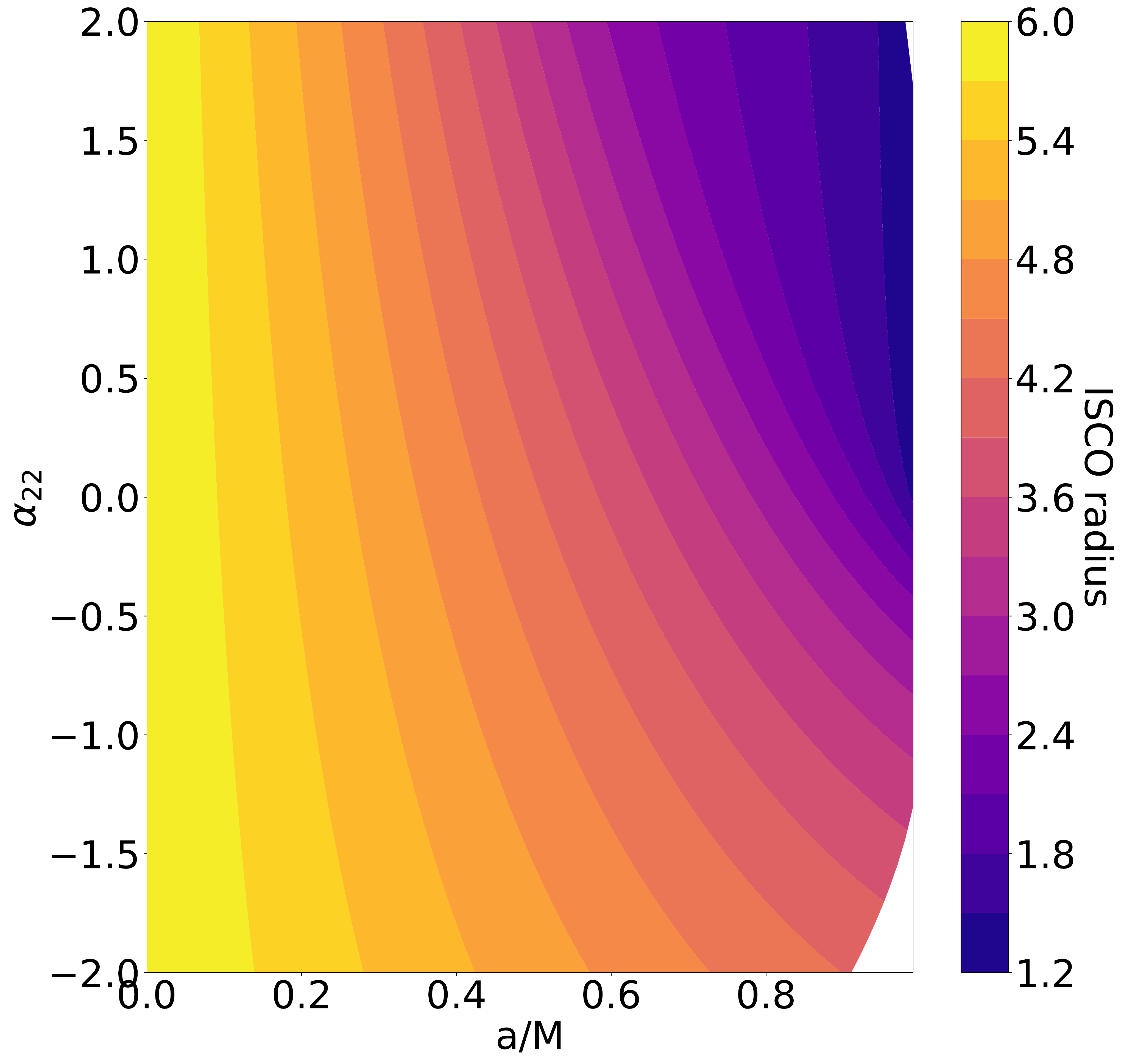}
\includegraphics[width=0.33\columnwidth{},clip=true]{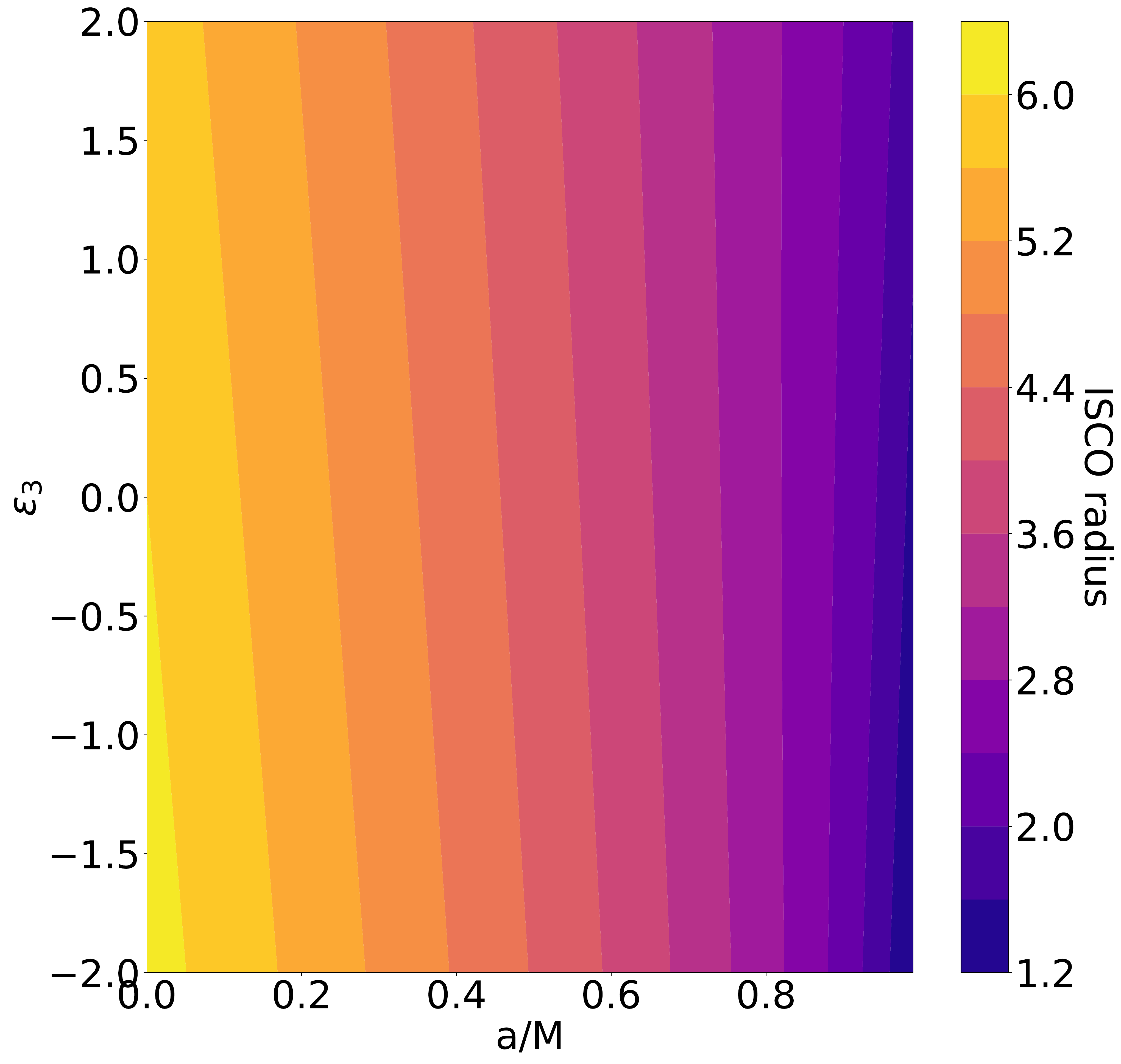}
\caption{ISCO radius as a function of spin and deformation parameter in the Johannsen spacetime. Only one deformation parameter is non-zero in each plot ($\alpha_{13}$, $\alpha_{22}$, and $\epsilon_{3}$, from left to right). The white regions on the plot are those that are excluded to avoid pathologies. Note that for the $\alpha_{52}$ case the ISCO radius is independent of $\alpha_{52}$ and matches that of Kerr.\label{fig:isco}}
\end{figure*}

The ISCO radius for the Kerr spacetime is
\begin{equation}
    r_{\text{ISCO}} = M\left\{3+Z_{2}\mp\left[\left(3-Z_{1}\right)\left(3+Z_{1}+2Z_{2}\right)\right]^{1/2}\right\},
\end{equation}
with
\begin{align}
    Z_{1} = & 1+\left(1-\chi^{2}\right)^{1/3}\left[\left(1+\chi\right)^{1/3}+\left(1-\chi\right)^{1/3}\right],
    \\
    Z_{2} = & \left(3\chi^{2}+Z_{1}^{2}\right)^{1/2}.
\end{align}
For comparison we can calculate the ISCO radius for the Johannsen spacetime with some of the leading order deformation parameters. In the case of the $\alpha_{52}$ parameter the ISCO radius matches that of Kerr. Otherwise, there is not a nice expression for the ISCO and instead we plot the ISCO as a function of the spin and deformation parameter in Fig.~\ref{fig:isco}.
%

\subsection{Finitely-thin and Thick Disks}

The far majority of black hole-accretion disk models assume an infinitesimally-thin disk regardless of the accretion rate of the black hole. This is clearly an approximation and in reality it is expected that the thickness of the disk should increase as the accretion rate increases. In fact, it is possible that some supermassive black holes in active galactic nuclei with accretion rates around or above the Eddington limit have inner accretion disks with thicknesses greater than the size of the black hole or even the radial extent of the inner disk. The thickness of the disk will of course modify the expected X-ray observations and may introduce a source of significant modeling error into estimates of the black hole's and disk's parameters.

Several disk models since the Shakura-Sunyaev and Novikov-Thorne thin disk models have been proposed to include larger thicknesses, some analytic, some phenomenological, and some numerical (see~e.g.~\cite{Abramowicz:2011xu} for a review). Very little work has been done on studying these various models in the context of tests of the Kerr spacetime and General Relativity, but we will here summarize that work.

\begin{figure}[t]
\begin{center}
\includegraphics[width=11.5cm,trim={0.5cm 0cm 0cm 0cm},clip]{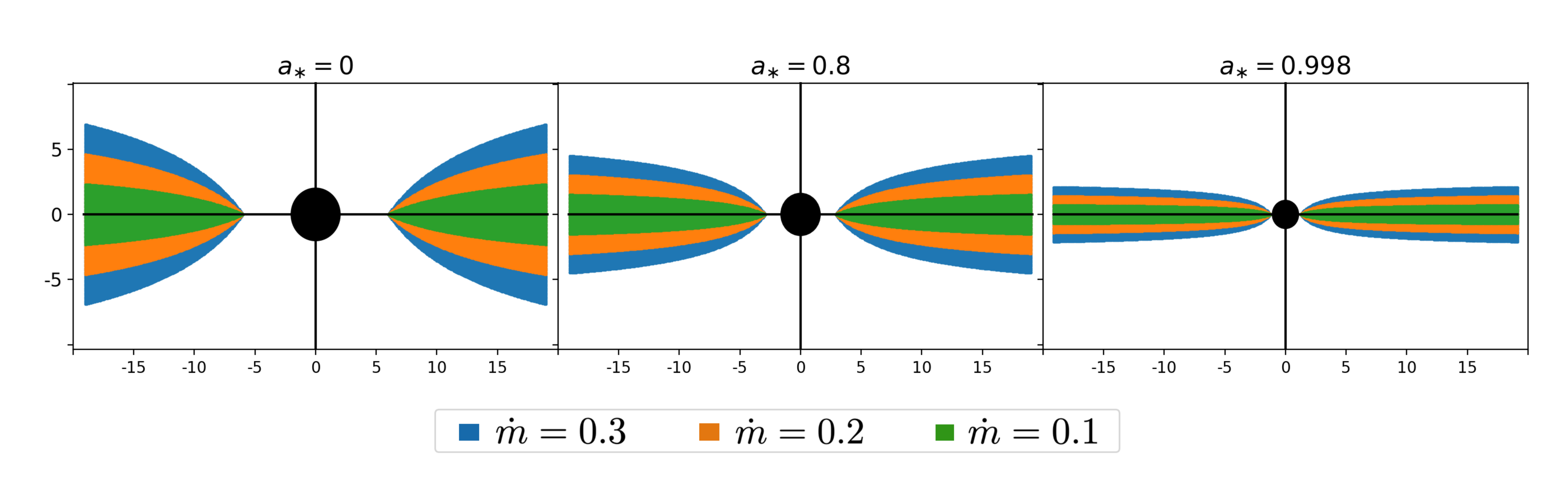}
\end{center}
\vspace{-0.5cm}
\caption{Profiles of accretion disks in Kerr spacetime for the black hole spin parameter $a_* = 0$, 0.8, and 0.998 and the Eddington-scaled mass accretion rate $\dot{M}/\dot{M}_{\text{Edd}} = 0.1$, 0.2, and 0.3. 
Figure from Ref.~\cite{Tripathi:2021wap}.
\label{f-finitelythin}}
\vspace{0.5cm}
\begin{center}
\includegraphics[width=8.5cm,trim={0.5cm 0cm 0cm 0cm},clip]{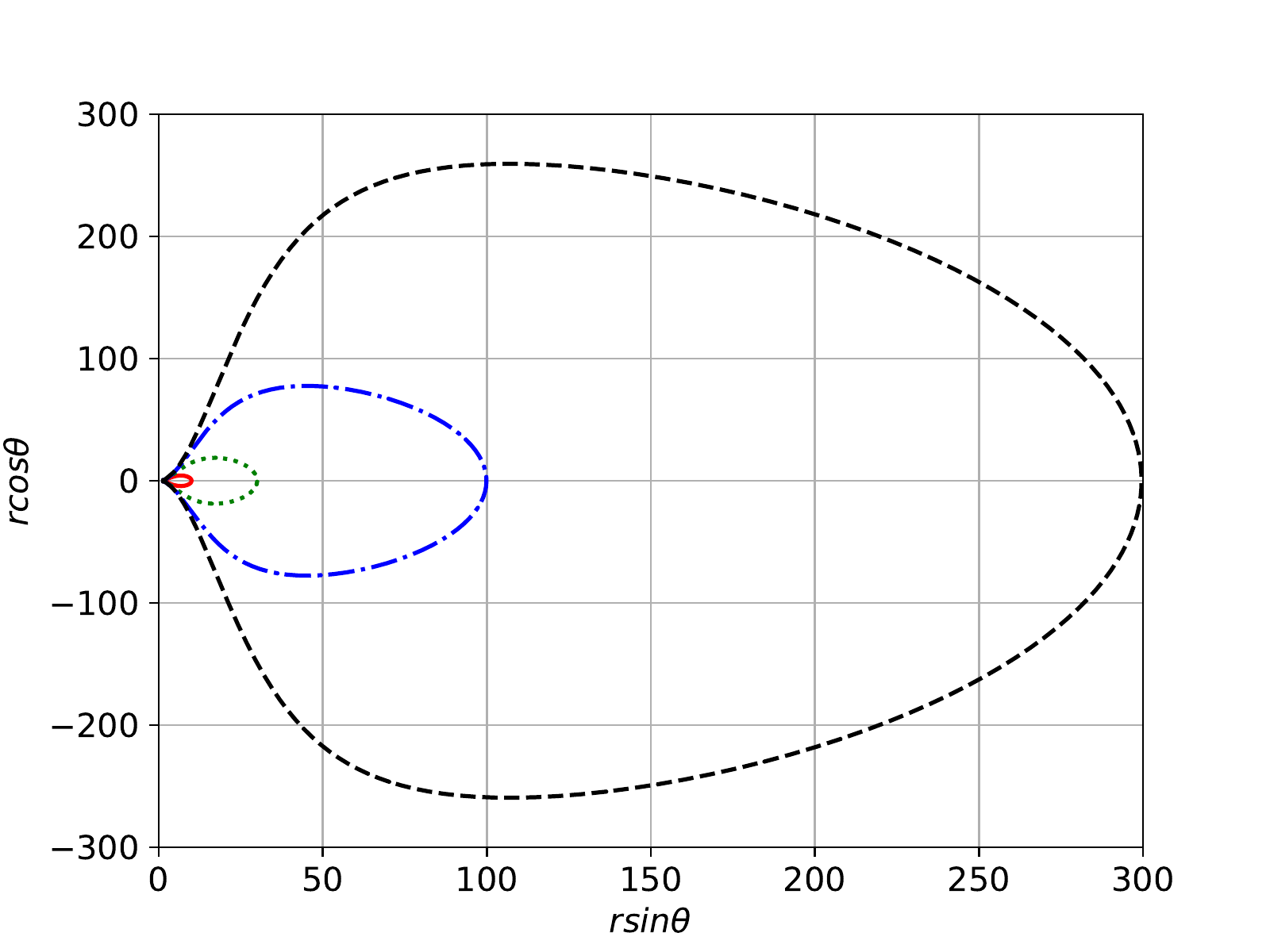}
\end{center}
\vspace{-0.2cm}
\caption{Examples of Polish donut disks in Kerr spacetime with spin
parameter $\chi = 0.95$. $r\sin\theta$ and $r\cos\theta$ in units $M = 1$. Figure from Ref.~\cite{Riaz:2019bkv}.
\label{f-donut}}
\end{figure}

There are two finite thickness disk models that have been studied in the context of tests of the Kerr spacetime. The first is a simple phenomenological extension of the infinitesimally-thin disk that introduces a height profile for the disk without modifying anything else~\cite{Taylor:2017jep}
\begin{equation}
    H=\frac{3}{2}\frac{1}{\eta}\left(\frac{\dot{M}}{\dot{M_{\text{Edd}}}}\right)\left[1-\left(\frac{r_{\text{ISCO}}}{r\sin\theta}\right)^{1/2}\right],
\end{equation}
where $\eta$ is the radiative efficiency of the disk, $r_{\text{ISCO}}$ is the ISCO radius, $\dot{M}$ is the accretion rate, and $\dot{M}_{\text{Edd}}$ is the Eddington accretion rate. The radiative efficiency is set to $\eta=1-E(r_{\text{ISCO}})$ and so the thickness of the disk for a given spacetime is controlled by the accretion rate ratio $\dot{M}/\dot{M}_{\text{Edd}}$ (see Fig.~\ref{f-finitelythin} for a profile of the disk). When using the infinitesimally-thin model to fit the finitely-thin disk model it was found that the spin can be significantly underestimated~\cite{Taylor:2017jep}. However, this bias seems to be dependent on the radiative efficiency (and in turn the spacetime properties) as a higher efficiency leads to a thinner disk. Depending on the black hole's spin, the disk thickness may play only a minor role and the infinitesimally-thin disk model is good enough with current data~\cite{Abdikamalov:2020oci,Tripathi:2021wap}. The same results are applicable to measuring departures from the Kerr solution.

The Polish donut model is an analytic model for perfect fluid in equilibrium orbiting around a black hole~\cite{1980AcA....30....1J, 1980A&A....88...23P, 1981AcA....31..283P, 1982ApJ...253..897P}. In essence, one can find equipressure surfaces of the fluid and these describe possible boundaries in the fluid. Under some choices, only one of these surfaces is physically realistic and can be chosen to be the thickness profile for the disk (see Fig.~\ref{f-donut} for a profile of the disk). There are a few studies on the Polish donut model in the context of testing the Kerr spacetime (see~e.g.~\cite{Li:2012ra, Riaz:2019bkv, Riaz:2019kat}). Generally, it is found that using a thin disk model when the reality is closer to a Polish donut model can introduce significant bias in the measurements of the spin and departures from the Kerr solution. This is particularly important in observations of active galactic nuclei where the accretion rates can be quite high and likely have disks closer to the Polish donut model than the thin disk.

\section{Observational Tests}
\label{sec:tests}

The X-ray radiation observed from black holes is thought to originate from accretion disks close to the objects. The accretion disks themselves, as described in the previous section, and the radiation emitted by the disks are influenced by the strong gravitational effects of the black hole spacetime. How the various X-ray observations are affected can be used to determine properties of the spacetime and help to test the Kerr hypothesis and General Relativity.

Performing tests on available data requires modeling the various observations in non-Kerr spacetimes, which in turn requires solving equations of motion for photons numerically. In the Kerr spacetime, as it contains the additional symmetry associated with the Carter constant, the equations of motion can be reduced to a pair of elliptic integrals~\cite{Chandrasekhar:1985kt,Bambi:2017khi} that are easily enough solved with a numerical integration scheme. Non-Kerr spacetimes do not necessarily have a Carter constant and the associated symmetry, and so require solving the full set of equations of motion consisting of two first-order and two second-order differential equations. In the following, the process for solving these equations, also known as ray-tracing, is summarized, before moving on to summarize the various X-ray observational tests.

The ray-tracing algorithm summarized here applies for any axially-symmetric stationary black holes and follows that described and used in~\cite{Psaltis:2010ww, Ayzenberg:2018jip, Abdikamalov:2020oci}. As the spacetime is stationary and axially-symmetric, it has a conserved energy $E$ and a conserved $z$-component of the angular momentum $L_{z}$. Using the definitions $E=-p_{t}$ and $L_{z}=p_{\phi}$, where $p^{\mu}$ is the photon's four-momentum, first-order differential equations for the $t$ and $\phi$ motion can be found
\begin{align}
    \frac{dt}{d\lambda'} & = -\frac{bg_{t\phi}+g_{\phi\phi}}{g_{tt}g_{\phi\phi}-g_{t\phi}^{2}},
    \\
    \frac{d\phi}{d\lambda'} & = \frac{g_{t\phi}+bg_{tt}}{g_{tt}g_{\phi\phi}-g_{t\phi}^{2}},
\end{align}
where $\lambda'\equiv E\lambda$ is the normalized affine parameter and $b\equiv L_{z}/E$ is still the impact parameter.

For the $r$ and $\theta$ motion the full second-order geodesic equations are used
\begin{align}
    \frac{dr}{d\lambda'} & = -\Gamma^{r}_{tt}\left(\frac{dt}{d\lambda'}\right)^{2}-\Gamma^{r}_{rr}\left(\frac{dr}{d\lambda'}\right)^{2}-\Gamma^{r}_{\theta\theta}\left(\frac{d\theta}{d\lambda'}\right)^{2}-\Gamma^{r}_{\phi\phi}\left(\frac{d\phi}{d\lambda'}\right)^{2}
    \nonumber \\
    & -2\Gamma^{r}_{t\phi}\left(\frac{dt}{d\lambda'}\right)\left(\frac{d\phi}{d\lambda'}\right)-2\Gamma^{r}_{r\theta}\left(\frac{dr}{d\lambda'}\right)\left(\frac{d\theta}{d\lambda'}\right),
    \\
    \frac{d\theta}{d\lambda'} & = -\Gamma^{\theta}_{tt}\left(\frac{dt}{d\lambda'}\right)^{2}-\Gamma^{\theta}_{rr}\left(\frac{dr}{d\lambda'}\right)^{2}-\Gamma^{\theta}_{\theta\theta}\left(\frac{d\theta}{d\lambda'}\right)^{2}-\Gamma^{\theta}_{\phi\phi}\left(\frac{d\phi}{d\lambda'}\right)^{2} 
    \nonumber \\
    & -2\Gamma^{\theta}_{t\phi}\left(\frac{dt}{d\lambda'}\right)\left(\frac{d\phi}{d\lambda'}\right)-2\Gamma^{\theta}_{r\theta}\left(\frac{dr}{d\lambda'}\right)\left(\frac{d\theta}{d\lambda'}\right),
\end{align}
where $\Gamma^{\alpha}_{\mu\nu}$ are the Christoffel symbols of the metric.

For simplicity, the reference frame and coordinate system are chosen such that the black hole is stationary at the origin and the black hole's spin angular momentum is along the $z$-axis. The mass of the black hole can also be set to unity, $M=1$, since the path the photon takes is only scaled by the mass.

When ray-tracing from the accretion disk to the observer, an observing screen is placed at a large enough distance $d$ such that gravitational effects are no longer important (e.g.~$d=10^6$). In reality the observer is much farther away, and so to simulate this photons are given final momenta that are perpendicular to the observing screen. This assures only those photon paths that would hit a screen at spatial infinity are considered. The final photon positions on the screen are scanned over, evolving the equations of motion backwards in time until the photon path encounters the disk, comes very near to the event horizon (in the coordinate systems commonly used the event horizon is a coordinate singularity), or returns to a relatively far distance from the black hole.

The initial position and four-momentum of each photon path in the Boyer-Lindquist coordinates of the black hole spacetime are~\cite{Bambi:2017khi}
\begin{align}
    r_{i} = & \left(d^{2}+\alpha^{2}+\beta^{2}\right)^{1/2},
    \\
    \theta_{i} = & \arccos\left(\frac{d\cos\iota+\beta\sin\iota}{r_{i}}\right),
    \\
    \phi_{i} = & \arctan\left(\frac{\alpha}{d\sin\iota-\beta\cos\iota}\right),
\end{align}
and
\begin{align}
    \left(\frac{dr}{d\lambda'}\right)_{i} = & \frac{d}{r_{i}},
    \\
    \left(\frac{d\theta}{d\lambda'}\right)_{i} = & \frac{-\cos\iota+d/r_{i}^{2}\left(d\cos\iota+\beta\sin\iota\right)}{\sqrt{r_{i}^{2}-\left(d\cos\iota+\beta\sin\iota\right)^{2}}},
    \\
    \left(\frac{d\phi}{d\lambda'}\right)_{i} = & \frac{-\alpha\sin\iota}{\alpha^{2}+\left(d\sin\iota-\beta\cos\iota\right)^{2}},
    \\
    \left(\frac{dt}{d\lambda'}\right)_{i} = & -\frac{g_{t\phi}}{g_{tt}}\left(\frac{d\phi}{d\lambda'}\right)_{i}-\left[\frac{g_{t\phi}^{2}}{g_{tt}^{2}}\left(\frac{d\phi}{d\lambda'}\right)_{i}^{2}\right.
    \nonumber \\
    & \left. -\left(g_{rr}\left(\frac{dr}{d\lambda'}\right)_{i}^{2}+g_{\theta\theta}\left(\frac{d\theta}{d\lambda'}\right)_{i}^{2}+g_{\phi\phi}\left(\frac{d\phi}{d\lambda'}\right)_{i}^{2}\right)\right]^{1/2},
\end{align}
where $\alpha$ and $\beta$ are the celestial coordinates on the observing screen, $\iota$ is the inclination angle between the observer's line-of-sight and the spin angular momentum of the black hole, and the component $\left(dt/d\lambda'\right)_{i}$ is computed by requiring the norm of the photon four-momentum to be zero. The impact parameter $b$ is calculated from the initial conditions and remains constant for each photon path as it is a conserved quantity.

If ray-tracing from a point-like source to the accretion disk, such as in the lamppost model (see Subsection~\ref{sec:reflect}), or from the disk back to the disk to account for higher-order reflection, the initial positions and momenta need to be modified and the equations of motion may be solved forwards in time depending on the physical setup.

With the basics of general relativistic ray-tracing introduced, we can now move on to summarizing the different black hole X-ray observations.

\subsection{Thermal Spectrum}

The thermal spectrum is the purely thermal radiation emitted by the disk. The Novikov-Thorne disk model provides time-averaged conservation equations for the radial structure of the disk, and these can be used to calculate the spectrum of thermal radiation~\cite{Page:1974he}. By mass conservation we have that
\begin{equation}
    \dot{M}=-2\pi\sqrt{-\textbf{g}}\Sigma(r)u^{r}=\text{constant},
\end{equation}
where $\dot{M}$ is the time-averaged mass accretion rate, $\textbf{g}$ is the metric determinant in the equatorial plane, $\Sigma(r)$ is the surface density of the disk, and $u^{r}$ is the radial four-velocity of the disk particles. From angular momentum and energy conservation we have an equation for the time-averaged radially-dependent energy flux $\mathcal{F}$
\begin{equation}
    \mathcal{F}=\frac{\dot{M}}{4\pi M^2}F(r),
\end{equation}
where $F(r)$ is a dimensionless function determined by the orbital characteristics of the disk
\begin{equation}
    f(r)=\frac{-\partial_{r}\Omega}{\left(E-\Omega L_{z}\right)^{2}}
    \frac{M^2}{\sqrt{-\textbf{g}}}
    \int^{r}_{r_{\text{in}}}\left(E-\Omega L_{z}\right)\left(\partial_{r'}L_{z}\right)dr'.
\end{equation}
$E$, $L_{z}$, and $\Omega$ are the radially-dependent energy, $z$-component of the angular momentum, and angular velocity, respectively, of the circular orbits in the equatorial plane, and $r_{\text{in}}$ is the innermost radius of the disk. In the Novikov-Thorne model the innermost radius of the disk is assumed to be the ISCO radius, however this is not necessarily the case in nature.

In place of the accretion rate $\dot{M}$, as it is not a directly observable parameter, one can use the Eddington ratio, $\ell=L_{\text{bol}}/L_{\text{Edd}}$, which is the ratio between the bolometric and Eddington luminosities, with $L_{\text{Edd}}\sim1.26\times 10^{38}(M/M_{\odot})$ erg/s. The accretion rate can then be rewritten in terms of the Eddington ratio and the Eddington accretion rate $\dot{M}_{\text{Edd}}$ as $\dot{M}=\ell\dot{M}_{\text{Edd}}$.

Furthermore, the radiative efficiency can be defined as $\eta\equiv L_{\text{Edd}}/\dot{M}_{\text{Edd}}$,~i.e.~the efficiency of conversion between rest-mass and electromagnetic energy of accreted particles. Assuming that the innermost radius of the disk is the ISCO radius, we can use that the energy radiated by a particle falling into a black hole is approximately equal to the binding energy at the ISCO radius~\cite{1973grav.book.....M}, and so the radiative efficiency is also given by $\eta=1-E(r_{\text{ISCO}})$. The accretion rate can now be written in terms of the Eddington ratio and the energy at the ISCO
\begin{equation}
    \dot{M}=\frac{\ell L_{\text{Edd}}}{1-E(r_{\text{ISCO}})}.
\end{equation}
The full expression for the radial energy flux is now
\begin{equation}\label{eq-thin-f-adim}
    \mathcal{F}(r)=\frac{\ell L_{\text{Edd}}}{4\pi\sqrt{-\textbf{g}}\left[1-E(r_{\text{ISCO}})\right]}\frac{-\partial_{r}\Omega}{(E-L_{z})^{2}}\int^{r}_{r_{\text{in}}}(E-\Omega L_{z})(\partial_{r'}L_{z})dr'.
\end{equation}

If the disk is in thermal equilibrium, the radiation emitted by the disk can be modeled as a black-body. By the Stefan-Boltzmann law, the radial energy flux is then related to the radial effective temperature of the disk by
\begin{equation}
    T_{\rm eff}(r)=\left(\frac{\mathcal{F}(r)}{\sigma}\right)^{1/4},
\end{equation}
where $\sigma\sim5.67\times10^{-5}$ erg/cm$^{2}$/s/K$^{4}$ is the Stefan-Boltzmann constant. Deviations from a perfect black-body spectrum, largely due to electron scattering in the disk atmosphere, can be taken into account by introducing the color correction term $f_{\rm col}$ and can be calculated by disk atmosphere models~\cite{Davis:2004jf,Davis:2006bk}. The color temperature is defined as $T_{\rm col} (r) = f_{\rm col} T_{\rm eff}$. In the rest-frame of the particles of the disk, the local specific intensity of the radiation emitted by the disk is (reintroducing the speed of light $c$)
\begin{equation}\label{eq-i-bb}
I_{\nu_e}(\nu_e) = \frac{2 h \nu^3_e}{c^2} \frac{1}{f_{\rm col}^4} 
\frac{\Upsilon}{\exp\left(\frac{h \nu_e}{k_{\rm B} T_{\rm col}}\right) - 1} \, ,
\end{equation}
where $h$ is the Planck constant and $k_{\rm B}$ is the Boltzmann constant. $\Upsilon$ is a function of the angle between the propagation direction of the photon emitted by the disk and the normal of the disk surface~\cite{Chandrasekhar:1985kt}. For isotropic emission, $\Upsilon =1$. The choice of $\Upsilon$ is an uncertainty in the model, but its impact is small.

Eventually, the photon flux number density as measured by a distant observer is~\cite{Bambi:2017khi}
\begin{eqnarray}\label{eq-cfm-n2}
N_{E_{\rm obs}} &=& A_1 \left(\frac{E_{\rm obs}}{\rm keV}\right)^2
\int \frac{1}{M^2} \frac{\Upsilon d\alpha d\beta}{\exp\left[\frac{A_2}{g F^{1/4}} 
\left(\frac{E_{\rm obs}}{\rm keV}\right)\right] - 1} \, ,
\end{eqnarray}
where $\alpha$ and $\beta$ are the Cartesian coordinates in the observer's sky, $F$ is the dimensionless function in Eq.~\ref{eq-thin-f-adim}, and $A_1$ and $A_2$ are two constants given by
\begin{eqnarray}
A_1 &=& \frac{0.07205}{f_{\rm col}^4} 
\left(\frac{M}{M_\odot}\right)^2 
\left(\frac{\rm kpc}{D}\right)^2 \, 
{\rm photons \,\, keV^{-1} \, cm^{-2} \, s^{-1}} \, , \\
A_2 &=& \frac{0.1331}{f_{\rm col}} 
\left(\frac{\rm 10^{18} \, g \, s^{-1}}{\dot{M}}\right)^{1/4}
\left(\frac{M}{M_\odot}\right)^{1/2} \, ,
\end{eqnarray}
where $D$ is the distance of the source from the observer.

The analysis of the thermal spectrum for measuring the physical parameters of the black hole-accretion disk system is known as the continuum-fitting method~\cite{Li:2004aq, McClintock:2013vwa, Zhang:1997dy}. The method is applied to stellar-mass black holes as the spectrum for those lies in the soft X-ray range, while it lies in the UV range for supermassive black holes and is observationally limited due to dust absorption. Additionally, the thermal spectrum is strongly dependent on the location of the inner edge of the disk as this sets the highest temperature. Generally, this is assumed to be at the ISCO radius and is a fair assumption when the X-ray binary source is in a high/soft state when the accretion luminosity is between 5\% and 30\% of the Eddington limit and the thermal spectrum dominates~\cite{Steiner:2010kd}.

The thermal spectrum has a fairly simple shape and is weakly dependent on the astrophysics of the disk making it great for extracting parameters of the black hole spacetime, however it has a significant drawback. Assuming a Kerr spacetime, the thermal spectrum depends on 5 parameters, the black hole mass, the black hole spin, the inclination angle, the accretion rate, and the distance to the source. For a non-Kerr background, at least one additional parameter needs to be introduced for the departure from Kerr. With all of these parameters and such a simple spectrum there are significant degeneracies when attempting to measure the parameters. For some systems the mass, inclination angle, and distance can be estimated from independent measurements, however that still leaves the black hole spin, accretion rate, and a possible deformation parameter. There are degeneracies between these three parameters and, depending on the specific deformation parameter, it may not be possible with only the analysis of the thermal spectrum to break them.

\begin{figure}[t]
\begin{center}
\includegraphics[width=5.8cm,trim={0.2cm 0cm 1.0cm 0cm},clip]{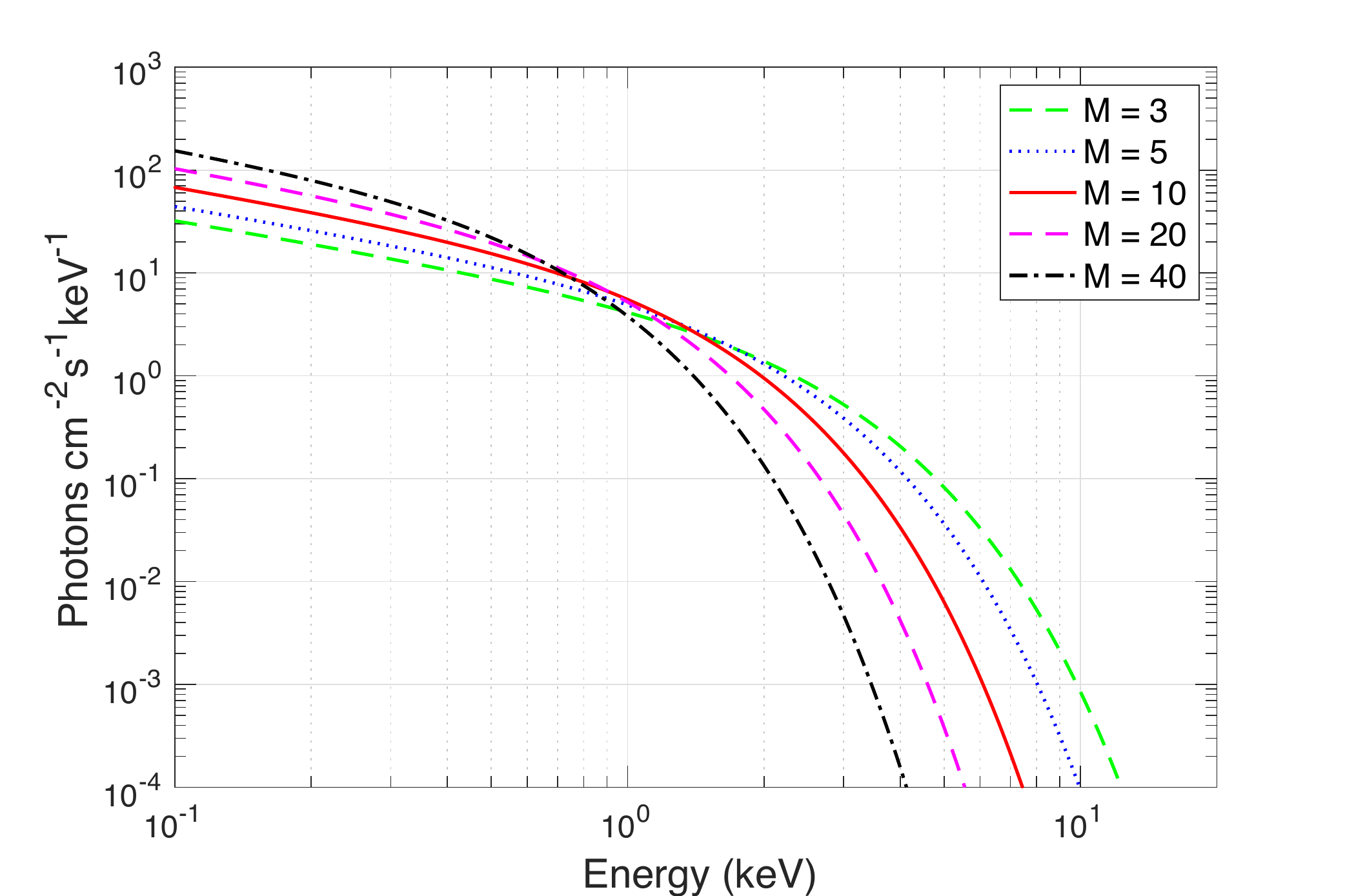}
\includegraphics[width=5.8cm,trim={0.2cm 0cm 1.0cm 0cm},clip]{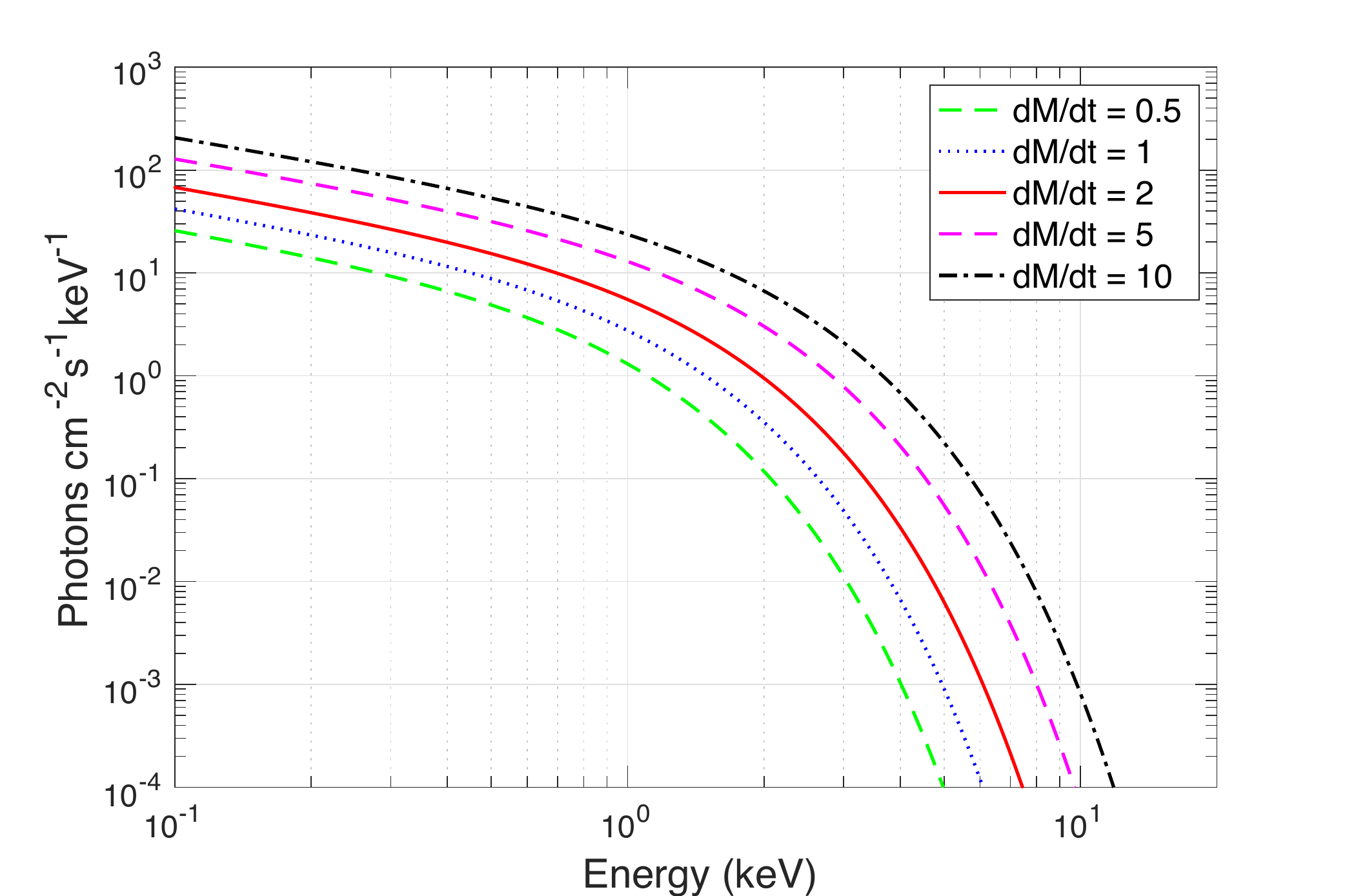} \\ \vspace{0.2cm}
\includegraphics[width=5.8cm,trim={0.2cm 0cm 1.0cm 0cm},clip]{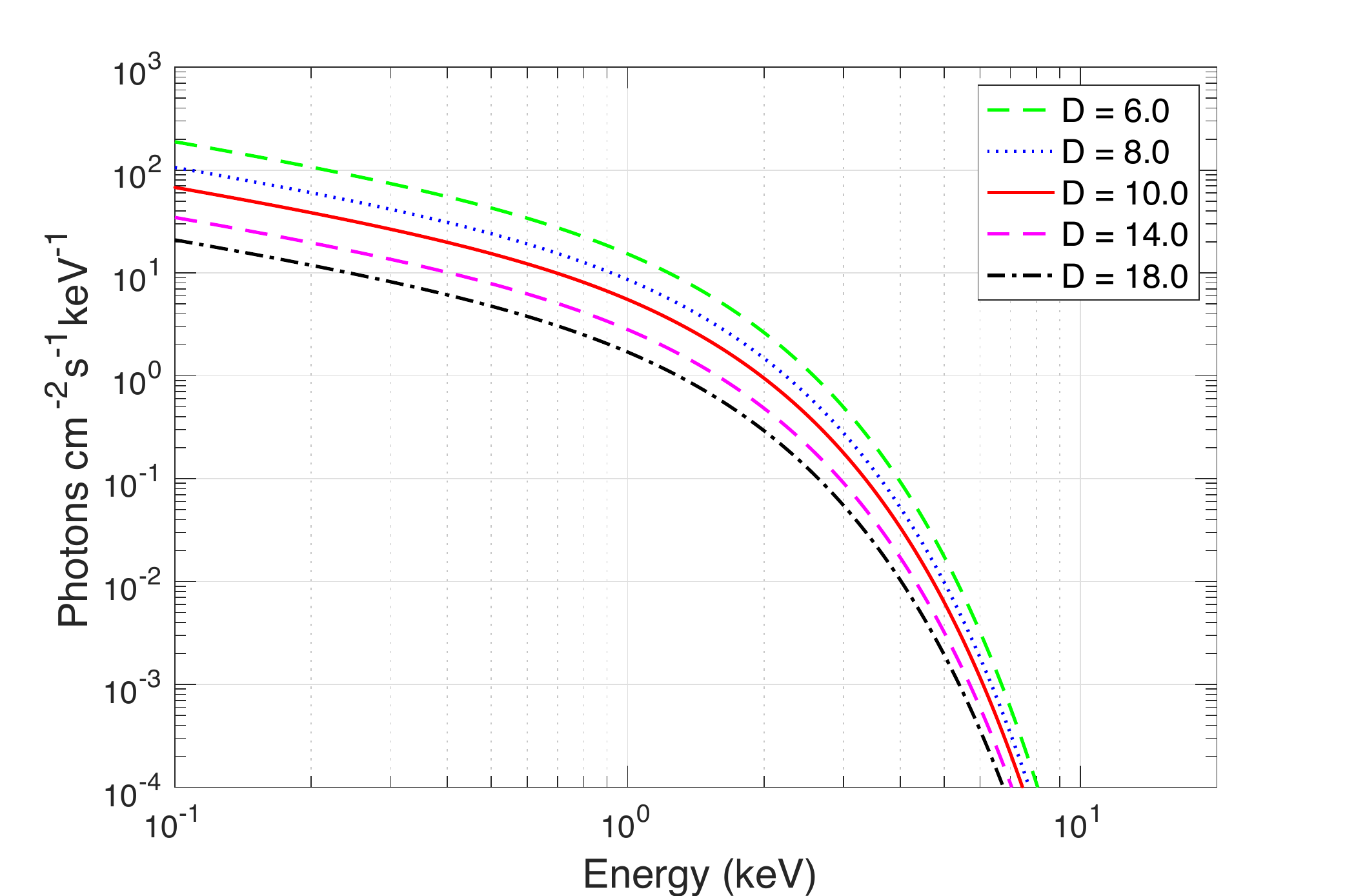}
\includegraphics[width=5.8cm,trim={0.2cm 0cm 1.0cm 0cm},clip]{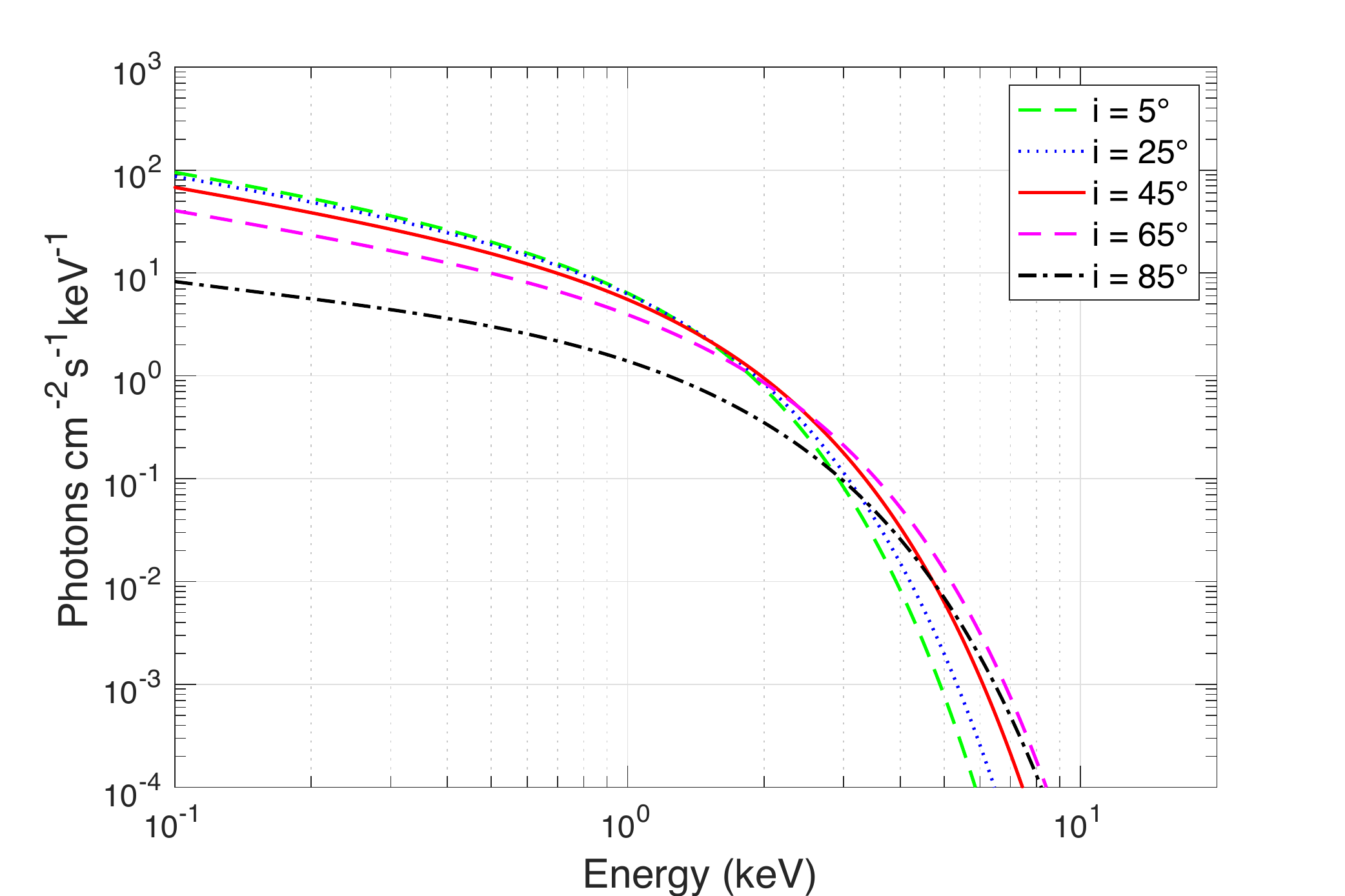} \\ \vspace{0.2cm}
\includegraphics[width=5.8cm,trim={0.2cm 0cm 1.0cm 0cm},clip]{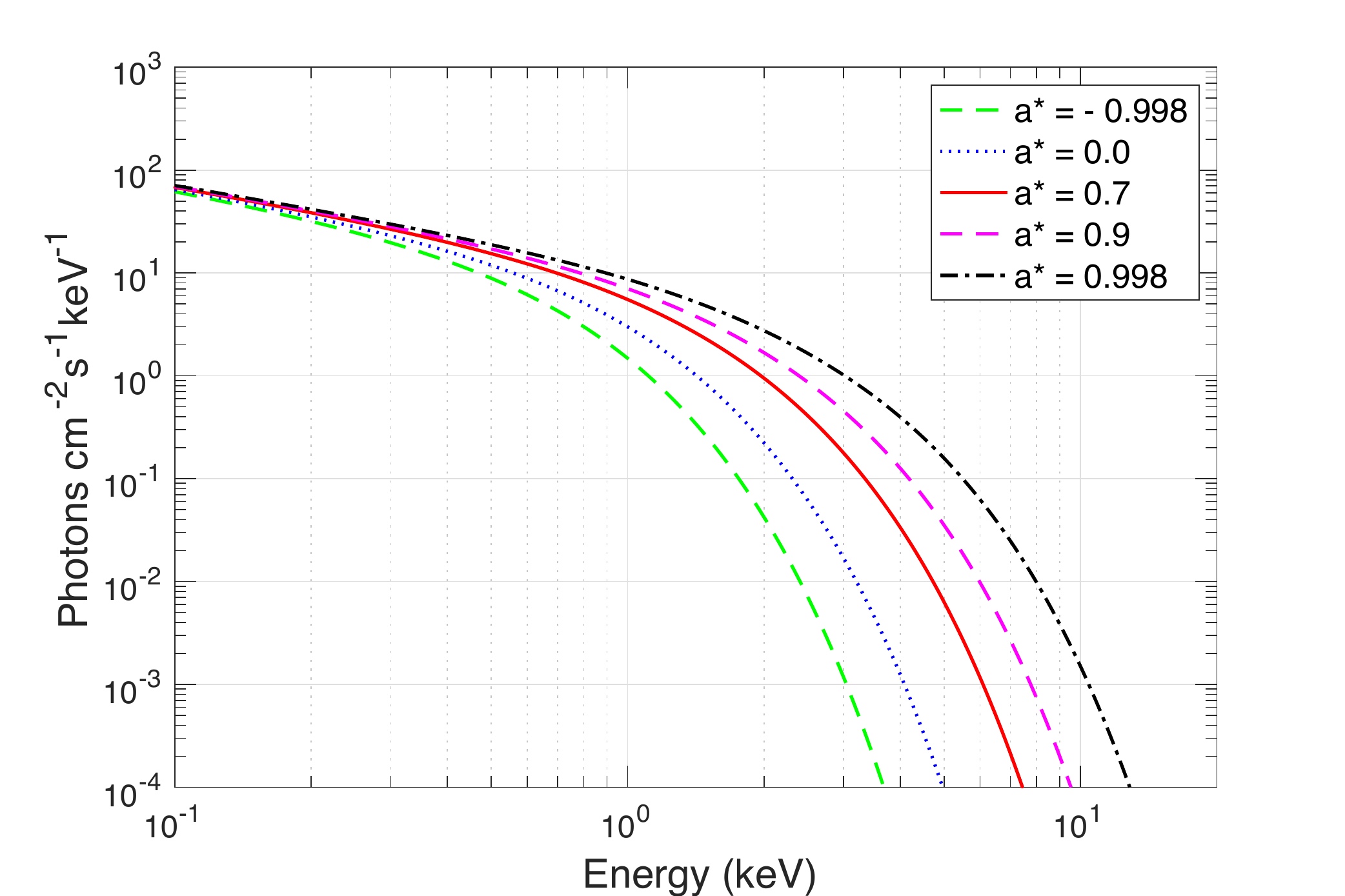}
\includegraphics[width=5.8cm,trim={0.2cm 0cm 1.0cm 0cm},clip]{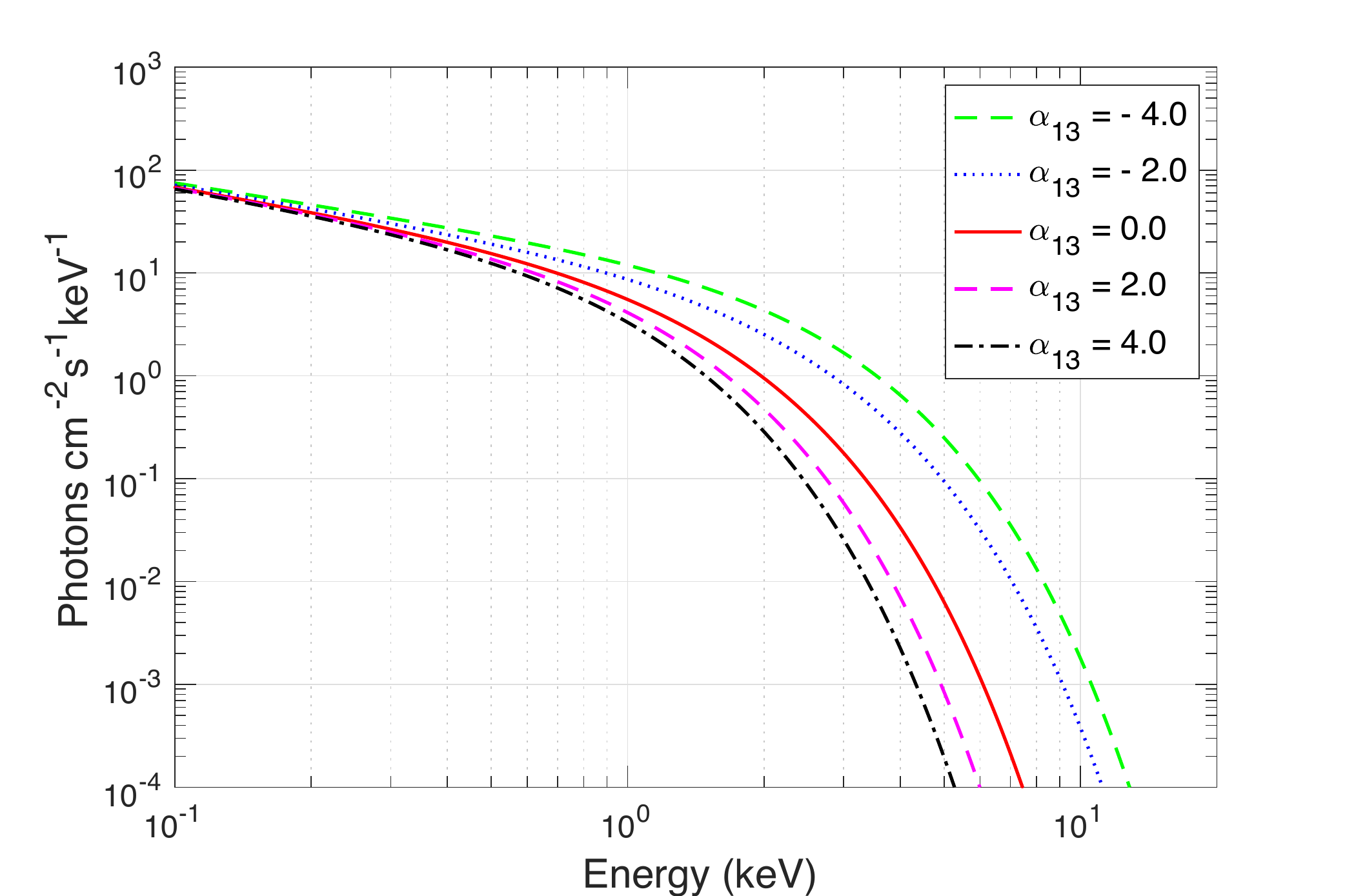}
\end{center}
\vspace{-0.2cm}
\caption{Thermal spectra of accretion disks as calculated by {\tt nkbb} for different values of the model parameters. $M$ in $M_\odot$, $\dot{M}$ in $10^{18}$~g/s, and $D$ in kpc. When not shown, the values of the model parameters are: $M = 10 \, M_\odot$, $\dot{M} = 2 \cdot 10^{18}$~g/s, $D = 10$~kpc, $i = 45^\circ$, $a_* = 0.7$, $\alpha_{13} = 0$. All simulations assume $f_{\rm col} = \Upsilon = 1$. Figures from Ref.~\cite{Zhou:2019fcg}. \label{f-para}}
\end{figure}

The thermal spectrum has been studied for non-Kerr spacetimes in a number of works (see~e.g.~\cite{Torres:2002td,Pun:2008ae, Pun:2008ua, Harko:2009xf, Harko:2009gc, Harko:2009kj, Bambi:2011jq, Bambi:2012tg, Kong:2014wha, Johannsen:2014arp, Ayzenberg:2017ufk}), and there is one thermal spectrum model particularly developed to be used with X-ray data analysis software known as \texttt{nkbb}~\cite{Zhou:2019fcg}. \texttt{nkbb} uses the techniques discussed throughout Sec.~\ref{sec:tests} to calculate the thermal spectrum for a given non-Kerr spacetime. Fig.~\ref{f-para} shows some spectra calculated by \texttt{nkbb} for different values of black hole mass, mass accretion rate, distance, inclination angle of the disk, spin parameter, and Johannsen deformation parameter $\alpha_{13}$. \texttt{nkbb} has been used to analyze X-ray data of some X-ray binaries and place constraints on departures from Kerr~\cite{Tripathi:2020qco, Tripathi:2020dni, Zhang:2021ymo, Tripathi:2021rqs}. Note that many of these studies analyzed the thermal spectrum and reflection spectrum of a source together in order to break the parameter degeneracies, and in general constraints were stronger through these joint analyses. A finite-thickness disk version of \texttt{nkbb} was recently created, but including this more realistic disk model seems to have minimal effect on the extracted parameters~\cite{Zhou:2020koa}.

\begin{figure}[t]
\begin{center}
\includegraphics[width=8.5cm,trim={0cm 0cm 0cm 0cm},clip]{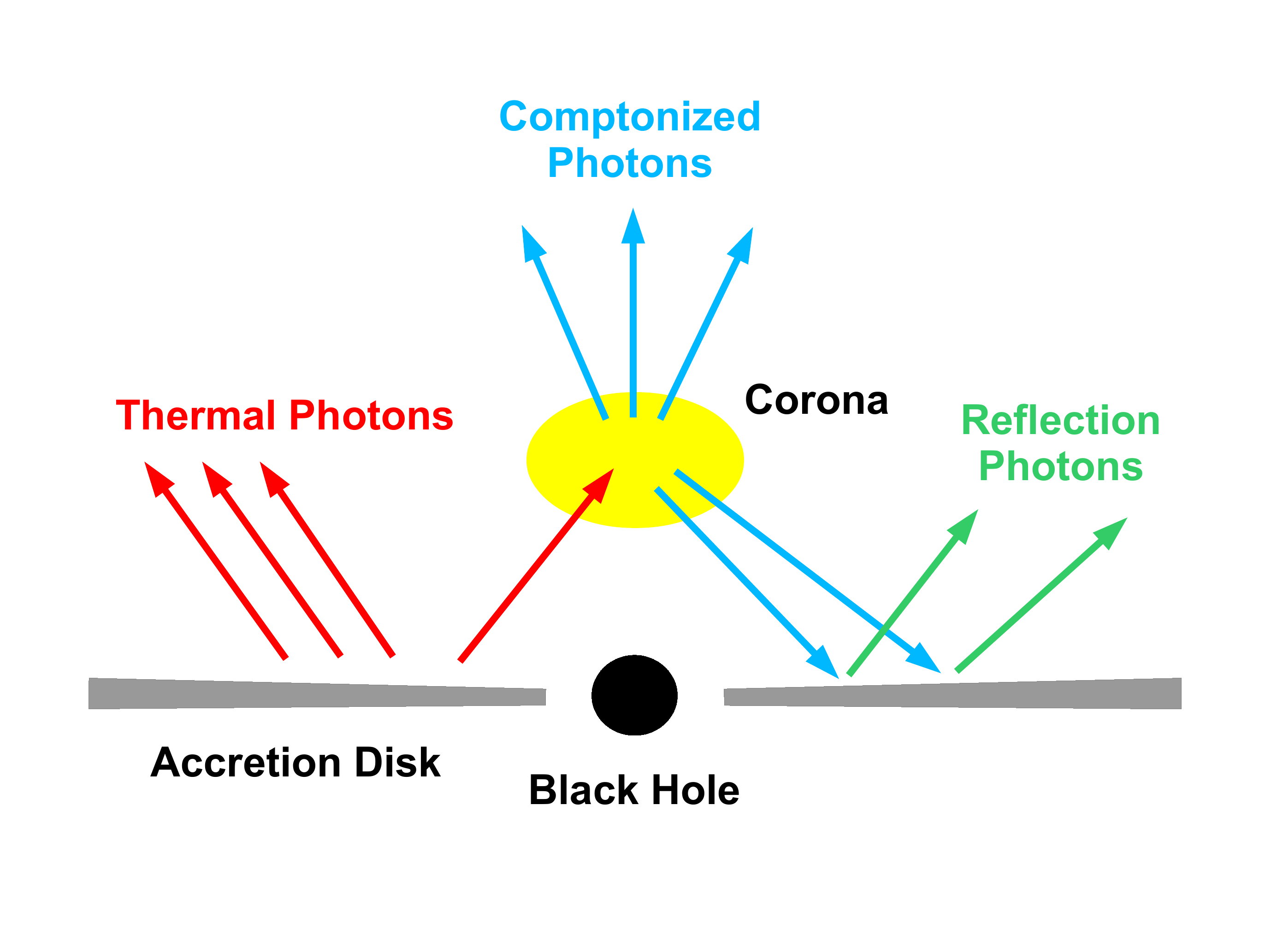}
\end{center}
\vspace{-1.0cm}
\caption{Cartoon illustrating the disk-corona model. A black hole is accreting from a geometrically thin and optically thick accretion disk. Thermal photons from the disk inverse Compton scatter off free electrons in the corona. A fraction of the Comptonized photons illuminate the disk and generate the reflection component.
Figure from Ref.~\cite{Bambi:2021chr}
\label{f-corona}}
\end{figure}

\subsection{Reflection Spectrum}
\label{sec:reflect}

The reflection spectrum is a component of the X-ray emission seen from both supermassive and stellar-mass black holes~\cite{Bambi:2020jpe}. The reflection spectrum depends on the presence of a relatively hotter cloud of gas in the black hole-disk system, known as a corona, see Fig.~\ref{f-corona}. The geometry of the corona is not currently known for certain, however several models have been proposed such as the lamppost geometry,~i.e.~a point-like cloud along the spin-axis of the black hole, and an atmosphere surrounding the accretion disk. This corona, regardless of its geometry, is believed to inverse-Compton scatter some of the thermal radiation originating from the disk, and produce a power-law spectrum. Some of this power-law spectrum returns to the disk and is then reprocessed and re-emitted, or reflected. The resulting reflection spectrum includes a forest of fluorescent emission lines. The most prominent of these emission lines is usually the K$\alpha$ iron line, which is at 6.4~keV in the case of neutral or weakly ionized iron and shifts up to 6.97~keV for H-like iron ions. In the rest-frame the K$\alpha$ iron line is fairly narrow, but as it travels to the observer from the disk it becomes broadened and skewed due to relativistic effects of the black hole spacetime and the motion of the disk. The shape of the observed line can thus be used to determine properties of the black hole spacetime. Additionally, the impinging radiation on the disk from the corona will also be influenced by the spacetime and so the overall reflection spectrum can also encode the black hole's properties.

\begin{figure}[t]
\begin{center}
\includegraphics[width=9.5cm,trim={0cm 0cm 0cm 0cm},clip]{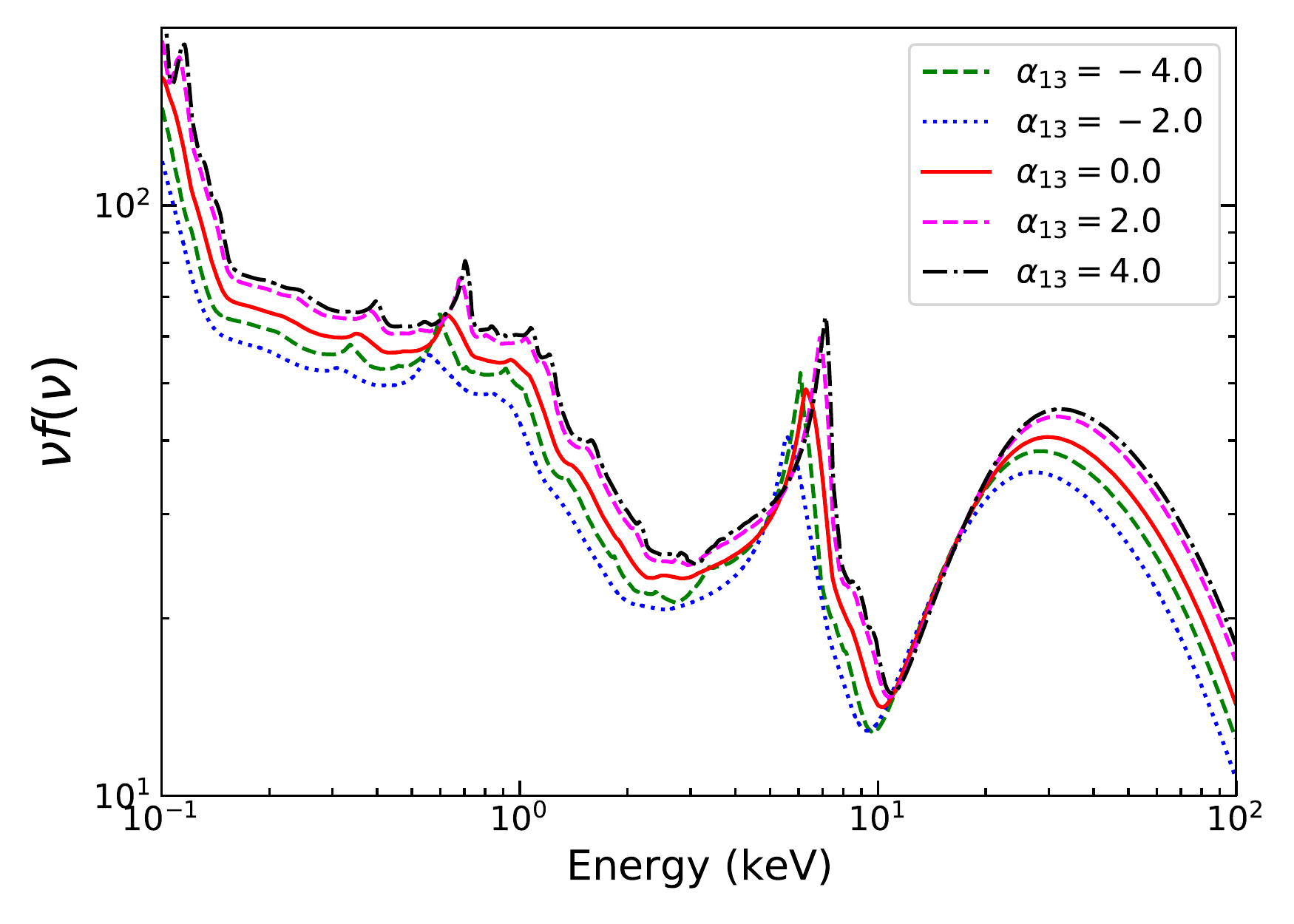} \\ \vspace{0.3cm}
\includegraphics[width=9.5cm,trim={0cm 0cm 0cm 0cm},clip]{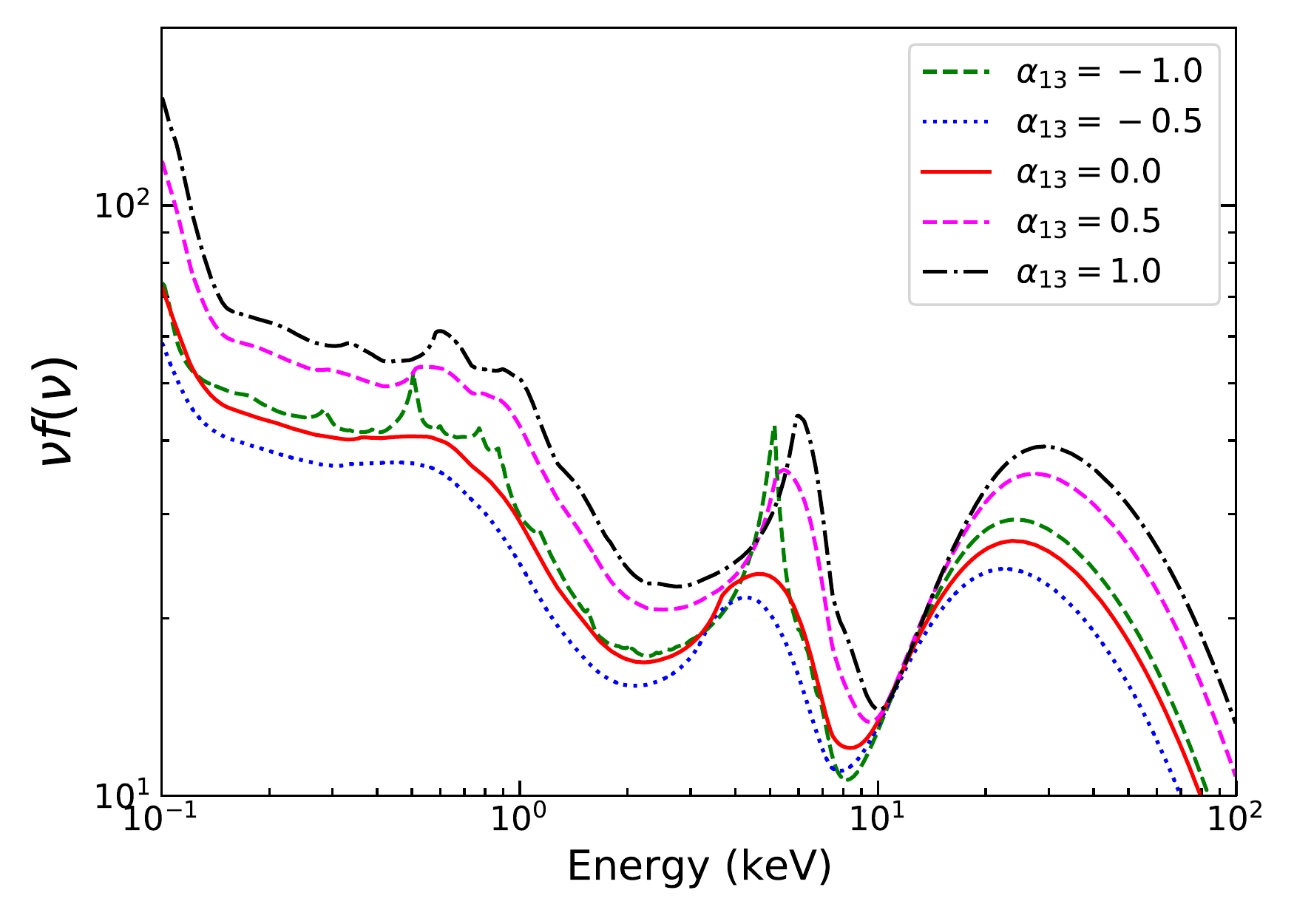}
\end{center}
\vspace{-0.3cm}
\caption{Synthetic reflection spectra of thin disks in the Johannsen spacetime for different values of the deformation parameter $\alpha_{13}$. These spectra are generated with {\tt relxill\_nk} assuming that the incident radiation has a power-law spectrum with photon index $\Gamma = 2$, the emissivity profile is described by a power-law with emissivity index $q = 7$, the ionization parameter of the disk is $\log\xi = 3.1$ ($\xi$ in erg~s~cm$^{-1}$), the disk has Solar iron abundance, the inclination angle of the disk is $i = 45^\circ$, and the black hole spin parameter is $\chi = 0.7$ (top panel) and 0.97 (bottom panel). 
Figure from Ref.~\cite{Bambi:2021chr}.
\label{f-relxillnk}}
\end{figure}

Calculating the reflection spectrum can be split into three parts: (1) Ray-tracing from the corona to the disk, (2) calculating the reprocessing inside the disk, and (3) ray-tracing from the disk to the observer. In simple models (1) is skipped and the impinging radiation on the disk is modeled as a power-law with photon index $\Gamma$ and a power-law emissivity profile, i.e.~the intensity on the disk $I\propto 1/r^{q}$, where $q$ is the emissivity index. The reprocessing inside the disk is usually handled through a separate model that can be incorporated with the overall model. For example, in \texttt{relxill}~\cite{Dauser:2013xv, Garcia:2013lxa} and \texttt{relxill\_nk}~\cite{Bambi:2016sac,Abdikamalov:2019yrr}, arguably the most advanced models for calculating the reflection spectrum for Kerr and non-Kerr black holes, respectively, the disk reprocessing is done with the model \texttt{xillver}~\cite{Garcia:2010iz, Garcia:2013oma}, which is incorporated directly into the models. The ray-tracing for non-Kerr spacetimes can be done following the procedure laid out in Secs.~\ref{sec:thindisks} and~\ref{sec:tests}. Examples of reflection spectra in the Johannsen spacetime generated by {\tt relxill\_nk} are shown in Fig.~\ref{f-relxillnk}.

When analyzing data, such as through the standard X-ray data analysis software \texttt{xspec}~\cite{1996ASPC..101...17A}, spectra must be calculated for a large number of values of the various physical parameters in the model. Calculating the spectra from scratch is not feasible with current technology and so an interpolation technique is used. The interpolation makes use of the Cunningham transfer functions~\cite{1975ApJ...202..788C, 1995CoPhC..88..109S} that allow for the recording of spectra in an efficient manner. Note this method can and is also used for the thermal spectrum.

We are primarily interested in the observed specific intensity $I_{o}(\nu_{o})$ at frequency $\nu_{o}$. By integrating the local specific intensity emitted from the accretion disk $I_{\nu_{e}}(r_{e},\theta_{e})$ over the observing screen, the observed specific intensity can be calculated. Here $\nu_{e}$, $r_{e}$, and $\theta_{e}$ are the frequency, radius of emission, and emission angle, respectively, of emitted photons in the frame of the location in the disk where the photons were emitted. To simplify this integration, the disk can be projected onto a plane perpendicular to the line of sight, i.e.~the observer's sky.

The observer can be placed at spatial infinity, $r=+\infty$, at an inclination angle $\iota$. We once again use the celestial coordinates $(\alpha,\beta)$ on the screen, and in terms of the photon momentum they are
\begin{equation}
    \alpha = \lim_{r\rightarrow\infty}\frac{-rp^{(\phi)}}{p^{(t)}}, \quad \beta = \lim_{r\rightarrow\infty}\frac{rp^{(\theta)}}{p^{(t)}},
\end{equation}
where $p^{(\alpha)}$ are the components of the photon's four-momentum with respect to a locally non-rotating reference frame~\cite{1972ApJ...178..347B}. These can be related to the usual photon four-momentum through a coordinate transformation, e.g.~$p^{\phi} = p^{(\phi)}/\sin\iota$. The solid angle on the observer's sky $d\omega$ can be related to the celestial coordinates, $d\alpha d\beta=D^{2}d\omega$, where $D$ is the distance between the observer and the black hole~\cite{1975ApJ...202..788C}.

Liouville's theorem~\cite{1966AnPhy..37..487L} in this scenario states that the ratio of the intensity and the frequency cubed is a constant, $I_{\nu}/\nu^{3}=\text{const.}$. This allows us to write the observed flux in terms of the emitted intensity
\begin{equation}
    F_{o}\left(\nu_{o}\right) = \int g^{3} I_{\nu_{e}}\left(r_{e},\theta_{e}\right)d\alpha d\beta,
\end{equation}
where $g$ is the redshift factor given in Eq.~\ref{eq:red}.

We can define the maximum and minimum frequency ratio $g^{*}$ at a given radius of the disk as
\begin{equation}
    g^{*} = \frac{g-g_{\text{min}}}{g_{\text{max}}-g_{\text{min}}} \in \left[0,1\right],
\end{equation}
where $g_{\text{min}}=g_{\text{min}}\left(r_{e},\iota\right)$ and $g_{\text{max}}=g_{\text{max}}\left(r_{e},\iota\right)$ are the minimum and maximum values, respectively, of the redshift factor $g$ for photons emitted at radius $r_{e}$ and detected by an observer with inclination angle $\iota$.

To carry out the integration over the accretion disk rather than the observer's sky, we perform a coordinate transformation from $\left(\alpha,\beta\right)$ to $\left(r_{e},g^{*}\right)$. This coordinate transformation makes use of the transfer function
\begin{equation}
    f\left(g^{*},r_{e},\iota\right) = \frac{1}{\pi r_{e}}g\sqrt{g^{*}\left(1-g^{*}\right)}\left\lvert\frac{\partial\left(\alpha,\beta\right)}{\partial\left(g^{*},r_{e}\right)}\right\rvert,
\end{equation}
where $|\partial\left(\alpha,\beta\right)/\partial\left(g^{*},r_{e}\right)|$ is the Jacobian.

Now the observed flux of the accretion disk can be written as
\begin{equation}
    F_{o}\left(\nu_{o}\right) = \int^{r_{\text{out}}}_{r_{\text{in}}}\int^{1}_{0}\frac{\pi r_{e}g^{2}f\left(g^{*},r_{e},\iota\right)}{\sqrt{g^{*}\left(1-g^{*}\right)}}I_{\nu_e}\left(r_{e},\theta_{e}\right)dg^{*}dr_{e},
\end{equation}
where $r_{\text{in}}$ and $r_{\text{out}}$ are the inner and outer radii of the disk, respectively.

For any given values of $r_{e}$ and $\iota$ the transfer function is generally a closed curve parameterized by $g^{*}$. There is one point for which $g^{*}=0$ and one point for which $g^{*}=1$, both in the accretion disk and, in turn, in the transfer function. Two curves connect these points, and so there are two branches of the transfer function, $f^{\left(1\right)}\left(g^{*},r_{e},\iota\right)$ and $f^{\left(2\right)}\left(g^{*},r_{e},\iota\right)$. The expression for the observed flux can then be split into two integrals
\begin{align}
    F_{o}\left(\nu_{o}\right) = & \int^{r_{\text{out}}}_{r_{\text{in}}}\int^{1}_{0}\frac{\pi r_{e}g^{2}f^{\left(1\right)}\left(g^{*},r_{e},\iota\right)}{\sqrt{g^{*}\left(1-g^{*}\right)}}I_{\nu_e}\left(r_{e},\theta_{e}^{\left(1\right)}\right)dg^{*}dr_{e}
    \nonumber \\
    & + \int^{r_{\text{out}}}_{r_{\text{in}}}\int^{1}_{0}\frac{\pi r_{e}g^{2}f^{\left(2\right)}\left(g^{*},r_{e},\iota\right)}{\sqrt{g^{*}\left(1-g^{*}\right)}}I_{\nu_e}\left(r_{e},\theta_{e}^{\left(2\right)}\right)dg^{*}dr_{e},
\end{align}
where $\theta_{e}^{\left(1\right)}$ and $\theta_{e}^{\left(2\right)}$ are the emission angles with relative redshift factor $g^{*}$ in branches 1 and 2, respectively.

The technique behind studying the reflection spectrum is known as X-ray reflection spectroscopy or, especially in the old literature, the iron line method. In contrast to the continuum-fitting method and as mentioned previously, X-ray reflection spectroscopy can be used for both stellar-mass and supermassive black holes. Additionally, it is currently the only method for determining the spins of supermassive black holes with any reliability.

Unlike the thermal spectrum, the reflection spectrum is fairly complex and more strongly dependent on the astrophysics of the accretion disk. However, the reflection spectrum is independent of the black hole's mass, the accretion rate, and the distance to the source, which in some instances can make it more suitable for studying black hole spacetimes. For a Kerr spacetime, the reflection spectrum depends on the black hole's spin, the inclination angle, the inner radius of the disk (which here is often usually assumed to be the ISCO radius, but is less important than for the thermal spectrum), the outer radius of the disk, the iron abundance, the ionization of the disk, and the emissivity profile of the disk. The emissivity profile depends on the assumptions that go into the model for the corona and can either be an ad-hoc profile such as a power-law or involve full ray-tracing from a chosen corona geometry. The uncertainties in the astrophysics of the disk can lead to significant degeneracies between the disk parameters and the parameters of the black hole spacetime, however the complexity of the reflection spectrum also makes it possible to break these degeneracies when the data is good enough.

The reflection spectrum is the most studied black hole X-ray observation, particularly in the context of tests of the Kerr solution and General Relativity (see~e.g.~\cite{Lu:2002vm, Torres:2003cc, Bambi:2012at, Johannsen:2012ng, Yang:2018wye}). Currently the only available model for testing the Kerr solution with X-ray reflection spectroscopy is \texttt{relxill\_nk}~\cite{Abdikamalov:2019yrr} and it has been used to study a number of black holes, both stellar-mass and supermassive, and place constraints on several non-Kerr spacetimes (see~e.g.~\cite{Tripathi:2019fms, Tripathi:2019bya, Zhou:2019kwb, Zhu:2020cfn, Tripathi:2021rwb, Abdikamalov:2021zwv}). As with the thermal spectrum and the \texttt{nkbb} model, \texttt{relxill\_nk} has a finite-thickness disk version, but analysis with this version does not seem to significantly impact the extracted parameters~\cite{Abdikamalov:2020oci, Tripathi:2021wap}.

\subsection{Other Tests}

The continuum-fitting and the iron-line methods are certainly the most popular and well-developed techniques for studying the strong gravity region with black hole X-ray data, particularly in the context of tests of the Kerr spacetime and General Relativity. However, there are other observational tests and we summarize the more well-studied ones here.

\subsubsection{X-ray Reverberation Mapping}

If the black hole-accretion disk system also contains a point-like corona situated along the black hole spin axis, as in the lamppost model, any flaring activity in the corona will lead to a time-dependent iron line~\cite{Reynolds:1998ie}. Different parts of the iron line originate from different parts of the accretion disk, e.g.~the redshifted tail is from the inner disk while the two peaks are from the receding and approaching wings of the disk. If the corona flares and our detector has a sufficiently large effective area and good time resolution, the light-propagation time to different parts of the disk manifests as a time-dependent brightening of different regions of the iron line spectrum. This is known as X-ray or iron line reverberation and can be used to study the properties of the black hole spacetime, i.e.~mapping.

Calculating the X-ray reverberation amounts to the same task as calculating the reflection spectrum described in Secs.~\ref{sec:tests} and~\ref{sec:reflect}, however the relative time-of-flight of the photons must be kept track of and this is usually not done in the case of the reflection spectrum. In the end, the transfer functions that are used to calculate the reflection spectrum have a time-dependence, are generally referred to as the 2D transfer functions, and these can be used to calculate the reverberation.

X-ray reverberation measurements are currently possible with supermassive black holes, but are most commonly used to study the properties of the disk and measure the mass of the black hole, rather than strong-field spacetime properties such as the black hole spin or deviations from the Kerr solution~\cite{Uttley:2014zca, Cackett:2021gad}. This is due to the low count rates from the iron line with current X-ray telescopes, such that reverberation measurements are not better than measurements of the standard time-integrated X-ray reflection for studying the black hole spacetime. Future telescopes with larger effective areas that lead to larger count rates should make reverberation mapping a more powerful technique than the standard reflection spectroscopy. The current poor knowledge of the coronal geometry is another problem that strongly limits the possibility of using X-ray reverberation mapping for precision measurements of the metric around a black hole. 

As reverberation mapping is a fairly new technique and is not particularly powerful with current telescopes, studies of using the technique to test the Kerr solution and gravity have focused only on the feasibility of reverberation mapping with current and future telescopes (see~e.g.~\cite{Jiang:2014loa, Jiang:2016bdj}). In the case of an ideal lamppost corona, these studies find that with current observing capabilities reverberation mapping does about as well as the time-averaged reflection spectrum for constraining the spin and departures from Kerr, while with future telescopes reverberation mapping will do significantly better (roughly an order of magnitude better constraints as compared with using the time-averaged iron line observation).

\subsubsection{Quasi-periodic oscillations}

\begin{figure}
\begin{center}
\includegraphics[width=9.0cm,trim={0cm 0cm 0cm 0cm},clip]{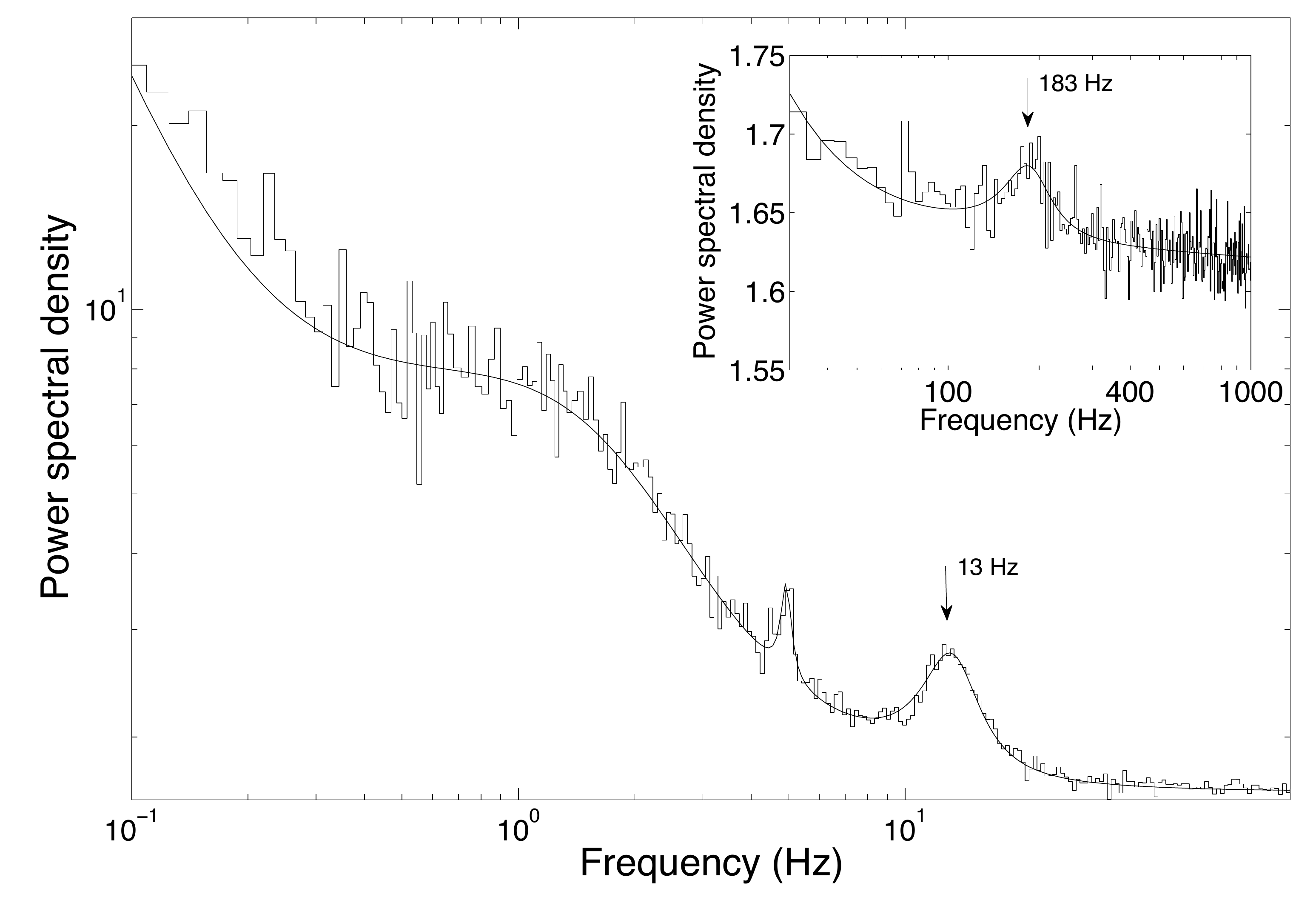}
\end{center}
\vspace{-0.2cm}
\caption{Power density spectrum from an observation of XTE~J1550--564 with a QPO around 5~Hz, another QPO at 13~Hz, and a third QPO at 183~Hz in the inset. Figure from Ref.~\cite{Motta:2013wwa}.
\label{f-qpos}}
\end{figure}

Quasi-periodic oscillations (QPOs) are a common, albeit not well understood, feature in the X-ray flux of stellar-mass black holes. They can also be observed in the spectra of intermediate-mass black hole candidates and supermassive black holes, but their detection is more challenging. QPOs are peaks in the X-ray power density spectrum of the source (see Fig.~\ref{f-qpos} for an example) and have been interpreted as some oscillation/motion in the disk, in the corona, or in the jet. Some of the proposed mechanisms are relativistic precession models~\cite{Perez:1996ti}, diskoseismology models~\cite{Motta:2013wga, Stella:1998mq}, resonance models~\cite{Abramowicz:2001bi, 2005A&A...436....1T}, and p-mode oscillations of accretion tori~\cite{Rezzolla:2003zx}. In most mechanisms, the QPO frequencies are related to the characteristic orbital frequencies of test-particles in the disk. These are the orbital or Keplerian frequency $\nu_{\phi}$, the radial epicyclic frequency $\nu_{r}$, and the vertical epicyclic frequency $\nu_{\theta}$. The orbital frequencies are fully determined by the background spacetime with no influence from the astrophysics of the disk. If QPOs are directly related to these frequencies analyzing them would be a powerful technique for studying the properties of black hole spacetimes. Indeed, QPOs can also be measured with high accuracy making them, in principle, more powerful than other techniques, assuming the correct mechanism can be determined.

Even without knowledge of the correct mechanism, there are a number of studies on using QPOs to test the Kerr spacetime; see~e.g.~\cite{Stuchlik:2008fy, Johannsen:2010bi, Aliev:2013jqz, Bambi:2012pa, Bambi:2013fea, Maselli:2014fca, Bambi:2016iip}. However, there is a further complication when using QPOs to study departures from Kerr due to the simplicity of the data. QPO data is essentially a pair (or rarely a trio) of frequencies where the X-ray power density peaks. The data being so simple is why it can be measured with high accuracy, but it also makes it difficult to break any degeneracies in the model. In general, the QPO frequencies depend on the mass, spin, and deformation parameter of the black hole spacetime, but these also tend to be strongly degenerate for a given pair of QPO frequencies. Thus, other independent measurements of the mass and spin of the black hole are required to break the degeneracies. Such is not necessarily the case with other X-ray observations like the reflection spectrum because these spectra are more complex and contain multiple independent features.

The drawbacks of the QPO observation can be exemplified in a couple of the aforementioned studies. In~\cite{Johannsen:2010bi}, the authors used a non-Kerr metric~\cite{Glampedakis:2005cf} to study whether the diskoseismology model and the 1:2 resonance model of the Keplerian and radial epicyclic frequencies could constrain the black hole spacetime parameters. The diskoseismology model can be used to constrain two parameters,~e.g.~the spin and deformation parameter, however an independent measurement of the mass would be required to break degeneracies. In the 1:2 resonance model, only one parameter can be constrained independently and so measurements of the mass and spin are required from other observations. Similarly, in~\cite{Bambi:2012pa}, a number of different resonance models were studied with the non-Kerr metric proposed in Ref.~\cite{Johannsen:2011dh}. As with the 1:2 resonance model, all of the resonance models studied could only measure one parameter of the spacetime. Additionally, when analyzing three systems with independent mass measurements, the resonance models did not match the spin estimates from thermal spectrum observations unless the deformation parameter was non-vanishing. It is not expected that these black holes are non-Kerr and rather that QPOs are still not well understood and require more study before they can be used to study black holes.

\subsubsection{X-ray Polarization}

The polarization of the thermal radiation from black hole accretion disks is an observation that could be possible with future X-ray polarimetric missions (as there are currently no X-ray polarimetric telescopes). The thermal radiation originating from a thin accretion disk is unpolarized at emission. However, as the radiation passes through the disk's atmosphere it will Thomson scatter off free electrons and become polarized at the level of a few percent. Since relativistic effects become more important closer to the black hole, the amount and angle of polarization will differ from the Newtonian prediction in the inner parts of the accretion disk. Polarimetric observations of the thermal spectrum could be used to measure the black hole spin and inclination angle~\cite{Li:2008zr, Schnittman:2009pm}, as well as test the Kerr metric~\cite{Krawczynski:2012ac, Liu:2015ibq} and the Weak Equivalence Principle. Degeneracies between the spin and deviations from General Relativity make it difficult to place constraints, but polarimetric measurements can still be useful in combination with other techniques.

\section{Conclusion}
\label{sec:concs}

\begin{figure}
\begin{center}
\includegraphics[width=12.0cm,trim={0cm 0cm 0cm 0cm},clip]{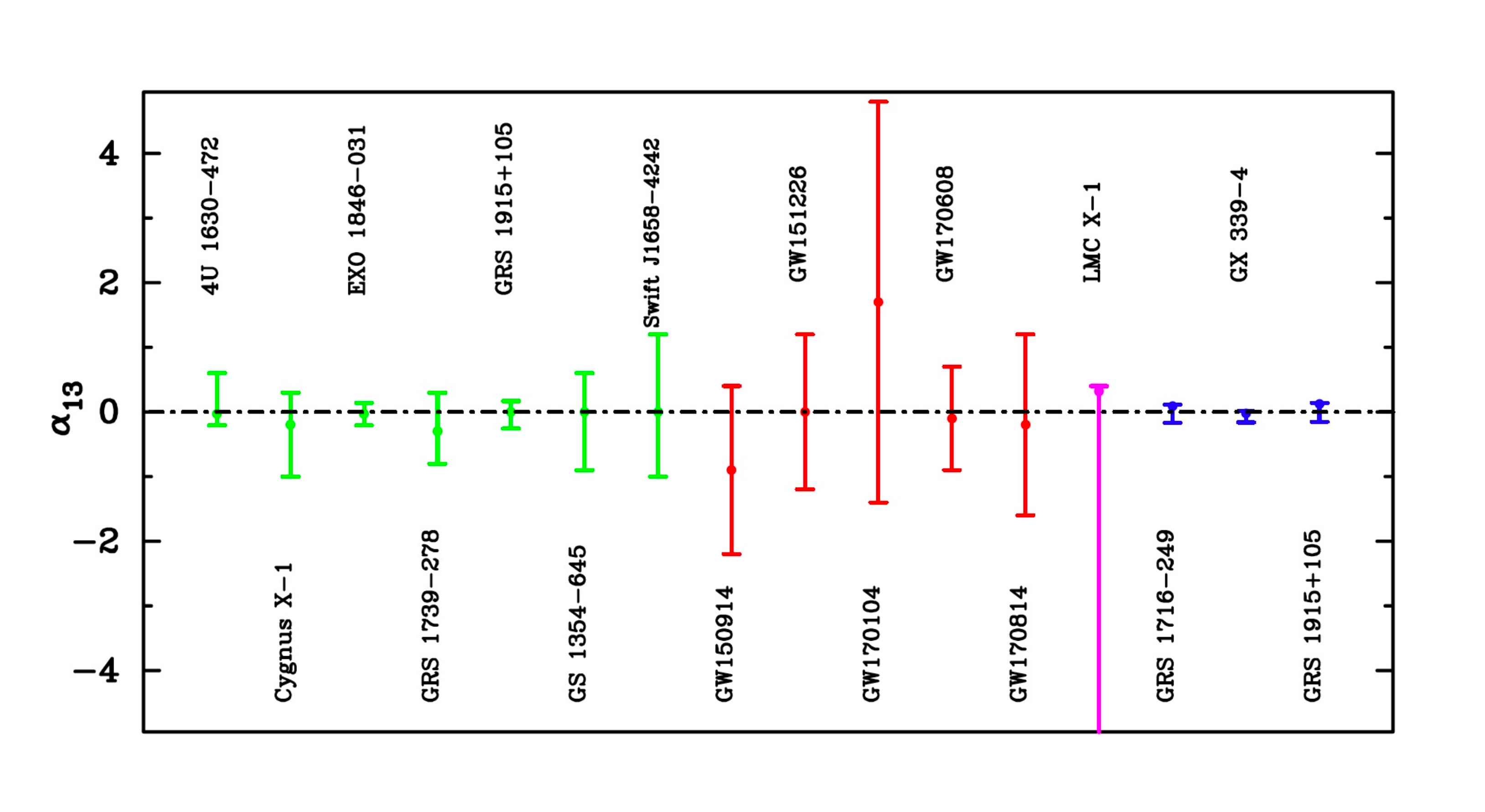}
\end{center}
\vspace{-0.5cm}
\caption{Summary of the 3-$\sigma$ constraints on the Johannsen deformation parameter $\alpha_{13}$ from stellar-mass black holes of Tab.~\ref{t-ssummary}. The constraints inferred from the analysis of reflection features are in green, those from gravitational wave data are in red, that from the continuum-fitting method in magenta, and the constraints obtained from the analysis of reflection features and the continuum-fitting method are in blue.
Figure from Ref.~\cite{Tripathi:2021rqs}
\label{f-summary}}
\end{figure}

Black hole X-ray observations can contribute significantly to our understanding of gravity and black holes, particularly in a regime not accessible to other tests of gravity. There has already been significant work in studying and developing models for the various observations, particularly the thermal spectrum and reflection spectrum, and these need to be and will be improved upon going forward. From the analysis of the thermal and reflection spectra of a number of stellar-mass and supermassive black holes, we have already quite robust constraints on the Kerr hypothesis, while little work has been done so far to test other predictions of General Relativity. Table~\ref{t-ssummary} and Fig.~\ref{f-summary} summarize current constraints on possible deviations from the Kerr metric in terms of the Johannsen deformation parameter $\alpha_{13}$ from stellar-mass black holes. The constraints are stronger when the continuum-fitting and the iron line methods are used together, but we note that for other deformation parameters the constraints from gravitational waves may be stronger than those from X-ray data. This is perfectly understandable: X-ray and gravitational wave observations measure different relativistic effects and therefore their constraining power depends on the specific deformation from the Kerr solution. 

A caveat, as already mentioned, with the constraints on possible deviations from the Kerr metric with black hole X-ray observations is the systematic bias due to the uncertainty in the accretion disk model. The currently used Novikov-Thorne thin disk model is certainly an approximation, but it is unclear to what level and in what ways this approximation introduces bias in not just constraints on deviations from Kerr, but even just the spin measurements when Kerr is assumed. As discussed, using an infinitesimally-thin disk model to fit a finitely-thin disk~\cite{Taylor:2017jep} or thick disk~\cite{Riaz:2019bkv, Riaz:2019kat} can lead to underestimates or overestimates of the spin, depending on the situation. However, the impact of using the wrong disk thickness when the black hole's spin is near maximal seems to be minimal on both spin estimates and constraining deviations from Kerr~\cite{Zhou:2020koa, Abdikamalov:2020oci, Tripathi:2021wap}. This needs to be better understood going forward.

Furthermore, the thickness of the disk is not the only simplification that is currently made. The coronal geometry that is used, if one is used at all, is simply a point source~\cite{Riaz:2020svt}. Secondary and higher order reflection is not considered, both in the reflection spectrum and thermal spectrum~\cite{Riaz:2020zqb}. Emission from the underside of the disk or that passes below the equatorial plane before going on to the observer is ignored~\cite{Zhou:2019dfw}. Even the astrophysics of the disk is currently simplified, such as the density profile~\cite{Abdikamalov:2021ues} or the ionization profile~\cite{Abdikamalov:2021rty}. All of these can introduce bias in estimates of the black hole spacetime's properties. They need to be studied, the biases accounted for, and eventually incorporated into more realistic disk models to reduce modeling bias as much as possible.

In the next years, with the launch of new X-ray telescopes such as ATHENA and eXTP, X-ray observations will become an integral part of studying black holes and gravity complementing observations of the black hole shadow, gravitational wave observations, and others. Precision tests of General Relativity with X-ray data will only be possible if we improve our theoretical models. The currently employed Novikov-Thorne accretion disk model will probably have to be replaced by simulation-based accretion disk models. A better understanding of the coronal geometry will also be required. It is important to work on the development of more accurate models prior to the launch of the new X-ray telescopes. The more accurate and precise data that is expected from the new telescopes will help us to better understand accretion disks and fine-tune the models that have been developed. With a more complete understanding of the disk physics, it will become possible to place quite stringent constraints on departures from the Kerr metric and General Relativity.

\begin{table}
\centering
{\renewcommand{\arraystretch}{1.3}
\begin{tabular}{lcccc}
\hline\hline
Source &  \hspace{0.5cm} Data \hspace{0.5cm}  & \hspace{0.5cm} $\alpha_{13}$ (3-$\sigma$) \hspace{0.5cm} & \hspace{0.5cm} Method \hspace{0.5cm} & Main Reference \\
\hline\hline
4U~1630--472 & \textsl{NuSTAR} & $-0.03_{-0.18}^{+0.63}$ & Fe-line & \cite{Tripathi:2020yts} \\
Cygnus~X-1 & \textsl{Suzaku} & $-0.2_{-0.8}^{+0.5}$ & Fe-line & \cite{Zhang:2020qbx} \\ 
EXO~1846--031 & \textsl{NuSTAR} & $-0.03_{-0.18}^{+0.17}$ & Fe-line & \cite{Tripathi:2020yts} \\
GRS~1716--249 & \textsl{NuSTAR}+\textsl{Swift} & $0.09_{-0.26}^{+0.02}$ & CFM + Fe-line & \cite{Zhang:2021ymo} \\
GRS~1739--278 & \textsl{NuSTAR} & $-0.3_{-0.5}^{+0.6}$ & Fe-line & \cite{Tripathi:2020yts} \\
GRS~1915+105 & \textsl{Suzaku} & $0.00_{-0.26}^{+0.17}$ & Fe-line & \cite{Zhang:2019ldz} \\
& \textsl{RXTE}+\textsl{Suzaku} & $0.12_{-0.27}^{+0.02}$ & CFM + Fe-line & \cite{Tripathi:2021rqs} \\
GS~1354--645 & \textsl{NuSTAR} & $0.0_{-0.9}^{+0.6}$ & Fe-line & \cite{Xu:2018lom,Tripathi:2020yts} \\
GW150914 & GWTC-1 & $-0.9 \pm 1.3$ & GW & \cite{Cardenas-Avendano:2019zxd} \\
GW151226 & GWTC-1 & $0.0 \pm 1.2$ & GW & \cite{Cardenas-Avendano:2019zxd} \\
GW170104 & GWTC-1 & $1.7 \pm 3.1$ & GW & \cite{Cardenas-Avendano:2019zxd} \\
GW170608 & GWTC-1 & $-0.1 \pm 0.8$ & GW & \cite{Cardenas-Avendano:2019zxd} \\
GW170814 & GWTC-1 & $-0.2 \pm 1.4$ & GW & \cite{Cardenas-Avendano:2019zxd} \\
GX~339--4 & \textsl{NuSTAR}+\textsl{Swift} & $-0.02_{-0.14}^{+0.03}$ & CFM + Fe-line & \cite{Tripathi:2020dni} \\
LMC~X-1 & \textsl{RXTE} & $< 0.4$ & CFM & \cite{Tripathi:2020qco} \\
Swift~J1658--4242 & \textsl{NuSTAR}+\textsl{Swift} & $0.0_{-1.0}^{+1.2}$ & Fe-line & \cite{Tripathi:2020yts} \\
\hline\hline
\end{tabular}}
\caption{Summary of the 3-$\sigma$ constraints on the Johannsen deformation parameter $\alpha_{13}$ from stellar-mass black holes with different techniques. CFM = continuum-fitting method; Fe-line = iron-line method; GW = gravitational waves (inspiral phase). Table from Ref.~\cite{Zhang:2021ymo}. \label{t-ssummary}}
\end{table}

\section{Cross-References}

\begin{itemize}
\item {\it Tests of Lorentz Invariance} (J. Wei and X. Wu)
\item {\it Fundamental physics with black holes} (J. Garcia)
\item {\it Fundamental physics with neutron stars} (J. Nattila and J. Kajava)
\end{itemize}

\end{document}